\documentclass{bmcart}

\usepackage{amsthm,amsmath}

\usepackage{multirow}
\usepackage{bbding}
\usepackage{graphicx}
\usepackage{color}

\startlocaldefs
\endlocaldefs

\begin{document}

\begin{frontmatter}

\begin{fmbox}
\dochead{Research}


\title{End-to-end Recording Device Identification Based on Deep Representation Learning}


\author[
addressref={aff1},                   
email={cyzeng@hbut.edu.cn}   
]{\inits{CZ}\fnm{Chunyan} \snm{Zeng}}
\author[
addressref={aff1},                   
email={840771073@qq.com}   
]{\inits{DL}\fnm{Dongliang} \snm{Zhu}}
\author[
addressref={aff2},
corref={aff2},
email={zfwang@mail.ccnu.edu.cn}
]{\inits{ZW}\fnm{Zhifeng} \snm{Wang}}
\author[
addressref={aff1},                   
email={wuxx1005@mail.hbut.edu.cn}   
]{\inits{MW}\fnm{Minghu} \snm{Wu}}
\author[
addressref={aff1},                   
email={xw@mail.hbut.edu.cn}   
]{\inits{WX}\fnm{Wei} \snm{Xiong}}
\author[
addressref={aff1},                   
email={nzhao@mail.hbut.edu.cn}   
]{\inits{NZ}\fnm{Nan} \snm{Zhao}}


\address[id=aff1]{
	\orgname{Hubei Key Laboratory for High-efficiency Utilization of Solar Energy and Operation Control of Energy Storage System, Hubei University of Technology}, 
	\street{Nanli Road 28},                     %
	\postcode{430068}                                
	\city{Wuhan},                              
	\cny{China}                                    
}
\address[id=aff2]{%
	\orgname{Department of Digital Media Technology, Central China Normal University},
	\street{Luoyu Road 152},
	\postcode{430079}
	\city{Wuhan},
	\cny{China}
}



\end{fmbox}


\begin{abstractbox}

\begin{abstract} 
Deep learning techniques have achieved specific results in recording device source identification. The recording device source features include spatial information and certain temporal information. However, most recording device source identification methods based on deep learning only use spatial representation learning from recording device source features, which cannot make full use of recording device source information. Therefore, in this paper, to fully explore the spatial information and temporal information of recording device source, we propose a new method for recording device source identification based on the fusion of spatial feature information and temporal feature information by using an end-to-end framework. From a feature perspective, we designed two kinds of networks to extract recording device source spatial and temporal information. Afterward, we use the attention mechanism to adaptively assign the weight of spatial information and temporal information to obtain fusion features. From a model perspective, our model uses an end-to-end framework to learn the deep representation from spatial feature and temporal feature and train using deep and shallow loss to joint optimize our network. This method is compared with our previous work and baseline system. The results show that the proposed method is better than our previous work and baseline system under general conditions.
\end{abstract}




\end{abstractbox}
%

\end{frontmatter}



\section*{1 Introduction}
Nowadays, the popularity of portable device sources makes it more and more convenient to obtain digital audio data \cite{Zeng2021a}. At the same time, the emergence of various powerful multimedia editing software also makes audio editing and modification easier \cite{Wang2022t}. In court and other important occasions, using tampered audio as electronic evidence will cause serious social problems \cite{1, 2, 35}. Therefore, it is of great significance to classify the source of digital audio data through recording device source identification technology \cite{3, 4}, and it has broad application prospects in judicial authentication and news information authenticity recognition \cite{Zeng2022a}.

Because the electronic components and structures in the audio capture device have different tolerances in nominal value, each audio capture device has a unique conversion function (i.e., frequency response) \cite{Zeng2020}. Therefore, each audio capture device will leave a unique internal trace in the voice recording  \cite{Zeng2021b}. The task of recording device source identification refers to identifying the class of device source from this inherent trace. This paper mainly focuses on portable recording devices (e.g., mobile phones, pad) source identification. In these previous works, most of the works used embedding methods to map devices into a feature space whose distance corresponds to the similarity of device sources. Although \cite{3, 5, 6, 7, 8, 9} had promoted the development of device source identification algorithms in the past few years, the task of device source identification is still a feature engineering task before the deep learning method came out.

Deep learning techniques can automatically extract highly abstract and complex features from raw data due to their powerful representation learning ability \cite{Zeng2022}. CNN (Convolutional Neural Networks) \cite{10,Zeng2022b} and DNN (Deep Neural Networks) \cite{11,Wang2022ac} are effective methods for device source identification, extracting the spatial correlation of related feature domains. LSTM (Long Short-Term Memory) can reasonably handle the temporal correlation in sequential data \cite{12,Lyu2022}, which has certain advantages in the sequence model with long-term memory. In addition, the deep learning methods directly make predictions based on the input data, which are conducive to the design of an end-to-end framework for device source identification tasks \cite{13,Zeng2021c,Wang2021}. Deep learning has many advantages, but device source identification based on deep learning still has many challenges. For example, in terms of feature representation learning, most device source identification tasks based on deep learning rely on extracting shallow spatial information from general acoustic features for identification. For instance, MFCC is used as an input to train a shallow CNN network for device source identification tasks \cite{Wang2021m,Wang2020h,Zeng2018,Wang2015b}. However, as far as we know, shallow structure networks have many limitations. With more and more data, limited samples and computing units will limit the network's generalization ability. In addition, these works only use DNN, CNN, and other networks to extract spatial information from device source features or directly use features that contain spatial information such as GSV. However, some device source features contain certain temporal information, such as MFCC. In terms of models, most traditional models or deep learning models contain many hand-designed parameters. The quality of feature selection depends on experience with human subjective factors. Moreover, some device source identification tasks lack transferability by combining deep learning methods with traditional machine learning methods.

Inspired by the success of deep learning for feature representation \cite{11,Zeng2020a,Wang2018a,Wang2017} and RNN for music classification tasks \cite{12,Wang2015a}, we propose a device source identification method based on the end-to-end framework for the fusion of spatial and temporal features. In this proposed method, first, to fully explore the spatial information and temporal information of device source, spatial and temporal feature extraction networks are built to extract spatial and temporal information from GSV feature and MFCC feature. Then, to fuse the two features information of the device source and increase the depth of the network, the attention mechanism is used to fuse spatial information and temporal information into a deep representation feature to obtain inherent traces left in the device source. Finally, to solve the problems of too many hand-designed parameters and poor transferability in the device source identification task, we build an end-to-end framework with deep-and-shallow loss to simplify the network and improve performance. 

This work is an extension of our previous work \cite{13}. In general, this work has the following contributions: 

(1) Parallel network representation learning is used to capture spatial and temporal information from device sources. 

(2) In terms of spatial and temporal information fusion, we add an attention mechanism that can assign the best weights to spatial and temporal features through autonomous network learning. 

(3) By establishing an end-to-end framework with strong transferability, a more compact model can be constructed, thereby reducing the complexity of the model. 

(4) In the optimization of the model, we design the deep-and-shallow loss to improve the model's generalization performance by assigning fixed weights to the deep and shallow losses, respectively.

The rest of this paper is organized as follows. Section 2 describes the related work. In section 3, we introduced the main methods of this paper. In section 4, we describe the parallel network used to extract the spatial information and temporal information of the device source in detail. The fifth part mainly introduces the attention mechanism and the back-end classification network. Section 6 designs experiments results and analysis. Section 7 introduces the conclusion and future work.

\section*{2 Related work}
In order to solve the problem of device source identification, many effective methods were proposed from different aspects. We divide the existing device source identification methods into two categories: feature-based methods and decision model-based methods. The feature-based method extracts the feature vector that characterizes the source information of the device from the original recording and then applies the existing vector data classification method. Decision model-based methods applied classifiers to solve the device source classification problems. The critical issue was to specify an appropriate learning function to measure the distance between two samples.

In this section, we discuss previous research work closely related to our method. In section 2.1, we summarize the existing feature-based device source identification methods. In section 2.2, we discuss several methods based on decision models.

\subsection*{2.1 The methods of device source identification based on features}
In the past, the features used in the device source identification method mainly included: (1)Devices source information characterization based on cepstrum features. (2)Representation of devices source information based on Gaussian Super Vector. (3)Devices source information based on deep features.
\subsubsection*{2.1.1 Devices source information characterization  based on cepstrum features}
Cepstrum features were widely used in the field of device source identification. Christian Kraetzer et al. \cite{14} proposed using Mel cepstrum features as machine fingerprints to identify device sources, which opened up the research field of device source identification. On this basis, Cemal Hanilçi et al. \cite{6} proposed the use of the MFCC (Mel Frequency Cepstral Coefficient) as a feature for device source identification. Since then, Qin Tianyun et al. \cite{15}, Ömer Eskidere et al. \cite{16}, Daniel Garcia-Romero et al. \cite{3}, Ling Zou et al. \cite{5}, Cemal Hanilçi et al. \cite{17} respectively verified the effectiveness of MFCC feature in device source identification tasks. Soon after, Aggarwal Rachit et al. \cite{18} first extracted the MFCC feature from the noise spectrum signal from the speech signal. This method is better than extracting the MFCC feature from speech segments based on the analysis of the experimental results. Meanwhile, other related cepstrum features are used in device source identification. For example, LFCC \cite{3}, BFCC (Bark Frequency Cepstrum Coefficient) and LPCC (Linear Prediction Cepstrum Coefficient) \cite{7,17}, PNCC (Power Normalized Cepstrum Coefficient) \cite{5} and improved PNCC \cite{19} proposed in the device source task. MFCC was calculated based on equally spaced frequency bands on the Mel scale instead of linear spaced frequency bands used. Chao Jin et al. \cite{20} used the principle of audio coding to quantize the spectral components to extract device source features by encoding Huffman below the masking threshold and achieved significant results.

\subsubsection*{2.1.2 Representation of devices source information based on Gaussian Super Vector}
GSV(Gaussian Super Vector) \cite{21}  was the feature data extracted from the mean vector of GMM (Gaussian Mixture Model). Traditional GSV included voice information and device source information. Yuechi Jiang et al. \cite{22} improved the quality of traditional GSV and mapped the traditional GSV to another dimension space. In this dimensional space, device source information and voice information can be divided into different dimensions, the back-end classifiers used SVM and SRC classifiers to conduct a comparative experiment. The experiment proved that the Kernel-Based GSV feature achieved better results compared to baseline experiments.

\subsubsection*{2.1.3 Devices source information based on deep features}
The essence of the deep neural network was to extract the inherent deep features of the data through the hidden layer of the network. There were two main ways to acquire deep features that supervised training or unsupervised training. In the feature extraction of supervised training, the device source data was used to train the appropriate network to extract the device source feature by giving the label. Unsupervised training extracted features by changing the data itself to extract feature data that could reflect the original. Inspired by this, Yanxiong Li et al. \cite{11,23} proposed two types of deep features: the first used MFCC features to build a deep neural network, and then extracted the output of the middle layer of the DNN network as the feature; the second used MFCC feature train a deep auto-encoding network and then used the output of the middle layer as the output feature. Experiments showed that the deep feature used by the author is better than the available feature. Xiaodan Lin et al. \cite{24} used the attention mechanism to assign feature weights to the frequency spectrum of different frequency bands. Experiments showed that the feature proposed by the author is more effective compared to the baseline.

\subsection*{2.2 The methods of device source identification based on decision model}

The decision model in device source identification mainly includes: (1) The device source identification decision model based on GMM (Gaussian Mixture Model). (2) The device source identification decision model based on SVM(Support Vector Machine). (3) The device source identification decision model based on the SRC (Sparse Representation Classifier). (4) The device source decision model based on deep learning model.

\subsubsection*{2.2.1 Device source identification decision model based on Gaussian Mixture Model}
The GMM-based classification model is to build a GMM for the data, compare the test data with the GMM model to calculate the probability, and finally take the highest probability. Such as Cemal Hanilçi et al. \cite{17} proposed to use the maximum amount of mutual information to measure the GMM. Comparative experimental results showed that better decision-making ability to use the maximum mutual information to train the GMM was better than the traditional training method in the case of short data.

\subsubsection*{2.2.2 Device source identification decision model based on Support Vector Machine}
The SVM classifier had perfect theoretical derivation and perfect interpretability in mathematics, so the SVM classifier was the most widely used model in machine learning. Different kernel functions were used in the SVM classifier to map features into high-dimensional space. Commonly used kernel functions were RBF (Radial Basis Function Kernel) and GLDS (Generalized Linear Discriminant Sequence Kernel) \cite{25}. Gianmarco Baldini et al. \cite{26,27} used FFT frequency features and compared the SVM classifier as a baseline classifier with the CNN network. Da Luo et al. \cite{28} proposed the BED feature of Fourier transform after signal framing. The back-end used SVM classification model still achieved positive results on 141 device sources.

\subsubsection*{2.2.3 Device source identification decision model based on Sparse Representation Classifier}
SRC used the internal elements of the dictionary as a basis function to transform the original feature data into 0, 1 sparse feature data by constructing a complete function dictionary. Ling Zou et al. \cite{29,30} used GSV to construct a database dictionary, and then used the K-SVD \cite{31} algorithm to calculate the score between the test device and the target device, at last, compared with a preset threshold to obtain the final recognition result. The K-SVD dictionary was obtained through unsupervised learning. Therefore, Ling Zou et al. \cite{29} proposed the use of D-KSVD (Discriminative K-SVD) \cite{32} algorithm to construct a supervised learning dictionary to improve the performance of devices source identification. When the training dataset was sufficient and complete, the SRC could improve the recognition efficiency by reducing the computational complexity. However, for classification problems with fewer samples, the improvement of SRC classification accuracy is a problem worthy of discussion.

\subsubsection*{2.2.4 Devices source identification decision model based on deep learning model}
The performance of deep learning models in various fields attracted attention. It could train not only large datasets but also has strong generalization and transferability. Gianmarco Baldini et al. \cite{8,26,27} used CNN in the back-end to surpass traditional classification methods and achieve better results. However, the fitting of a shallow network does not fully reflect the effect of deep learning. Further, Chunyan Zeng \cite{13} et al. used a multi-feature parallel convolution network, combined with the attention mechanism for device source identification to achieve an improved effect. With further research on device source identification, the feature dimension of device source identification and the number of devices for device source identification tasks will further increase. The general statistical model required a large amount of calculation and is challenging to meet these demands. The application of the deep model could undoubtedly overcome these shortcomings.

\section*{3 Methods}
We integrate the steps of device source identification into a single network in Fig. 1. The input of this network includes "training" speech and "testing" speech. The output is a single node indicating the category. We propose the deep-and-shallow loss to optimize this network. In section 3.1, we describe the overview architecture of our end-to-end device source identification framework. The details of its important components will be discussed in section 4 and section 5. In section 3.2, we describe the training process of the end-to-end framework.

\begin{figure}[!htbp]
\centering
\includegraphics[width=4.5in]{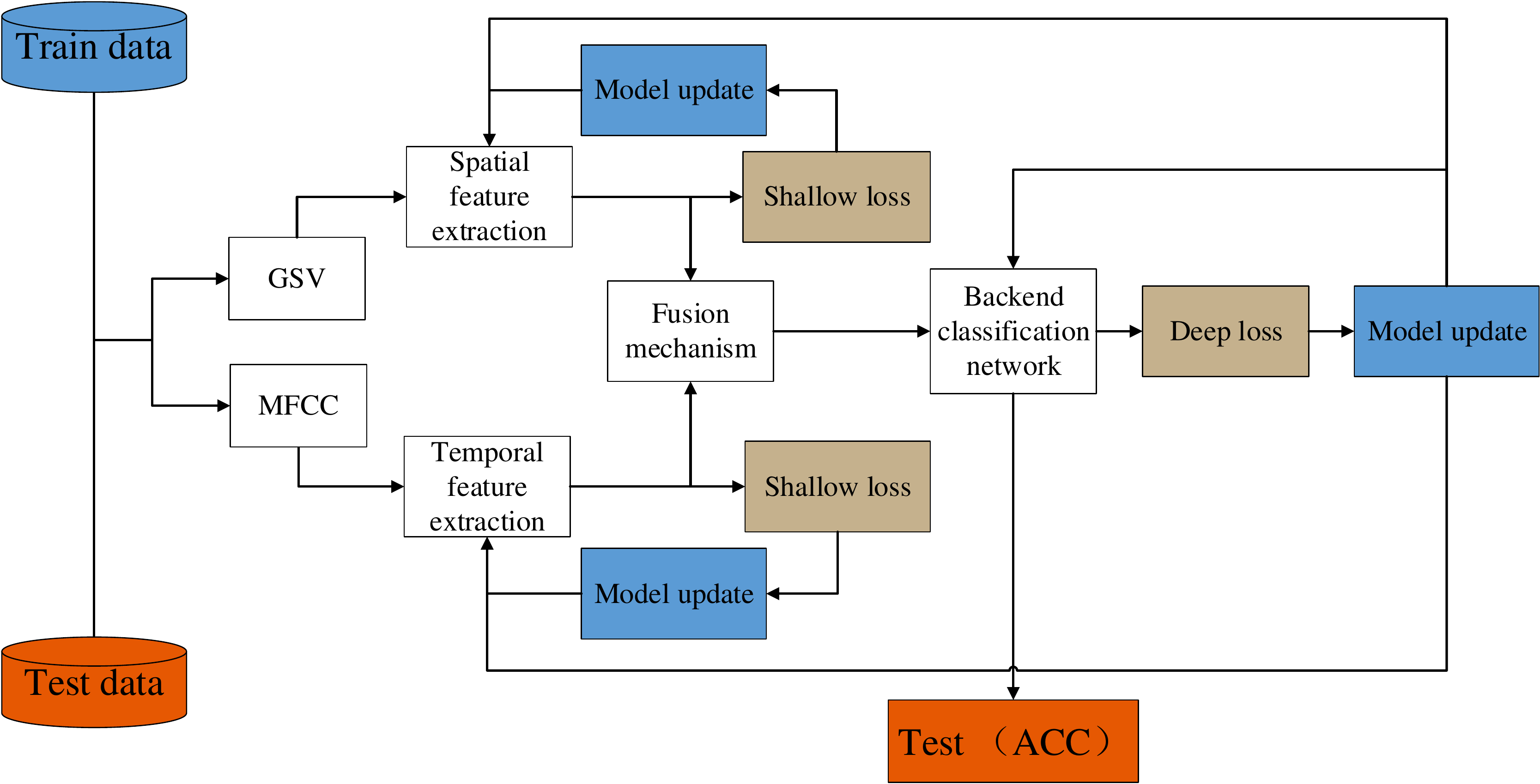}
  \caption{The architecture of our end-to-end framework.}
\end{figure}

\subsection*{3.1 End-to-end framework}
The structure of the end-to-end framework for device source identification is depicted in Fig. 1. The framework consists of a parallel feature extraction network module, a feature fusion module, a back-end classification module, and a deep-and-shallow loss module. In the parallel feature extraction module, to extract device source spatial information and temporal information, GSV and MFCC features are put into the spatial and temporal information extraction network, respectively. In the feature fusion module, we use the attention mechanism to fuse the device source spatial information and temporal information to obtain more representative features. Based on our previous work \cite{13}, we used full connection with a decreasing number of nodes in the back-end classification module in this paper. In the deep-and-shallow loss module, we designed a deep-and-shallow loss for the end-to-end framework, and the specific information is described in section 3.2.

\subsection*{3.2 End-to-end training}
As shown in Table 1, our end-to-end training process includes initialization, data loading, batch processing, loss calculation, gradient calculation, and parameter update. Steps 1 to 3 represent the initialization of the network and the loading of data, which will be described in section 6. Step 4 belongs to the forward propagation of network training, including calculating deep-and-shallow loss. Steps 5-9 belong to the back-propagation stage of the network, including calculation of iterative gradients and update of network parameters.

\newcommand{\tabincell}[2]{\begin{tabular}{@{}#1@{}}#2\end{tabular}}
\begin{table}[!htbp]
\centering
\caption{End-to-End joint training.}
    \begin{tabular}{l}
        
        \hline
        Optimization of end-to-end parameter update algorithm based on deep-and-shallow loss \\
        \hline
        dataset: 45 device sources, each device source is 514 sentences for training\\
        \hline
        1: initialized: $W_{t e m}$ ,$W_{s p a}$ ,$W_{att}$ ,$W_{class}$\\
        2:\hspace{0.3cm} for k=1,...,K (K=epoch) \\
        3:\hspace{0.6cm} for t=1,...,T (T=514/batch size) \\
        4:\hspace{0.9cm} deep-and-shallow loss:   $L(t)=\lambda_{1} L_{T}(t)+\lambda_{2} L_{S}(t)+\left(1-\lambda_{1}-\lambda_{2}\right) L_{A}(t)$ \\  
        5:\hspace{0.9cm} backpropagation error:$\frac{\partial L(t)}{\partial x_{i}(t)}$\\  
        6:\hspace{1cm}$W_{\mathrm{tem}}(t+1)$$\gets$$W_{\mathrm{tem}}(t)-\mu * \frac{\partial L_{T}(t)}{\partial W_{\mathrm{tem}}(t)}-\mu * \frac{\partial L_{A}(t)}{\partial W_{\mathrm{tem}}(t)}$ \\    
        7:\hspace{0.9cm} $W_{\mathrm{spa}}(t+1)$$\gets$$W_{\mathrm{spa}}(t)-\mu * \frac{\partial L_{s}(t)}{\partial W_{\mathrm{spa}}(t)}-\mu * \frac{\partial L_{A}(t)}{\partial W_{\mathrm{spa}}(t)}$  \\
        8:\hspace{0.9cm} $W_{\mathrm{att}}(t+1)$$\gets$$W_{\mathrm{att}}(t)-\mu * \frac{\partial L_{A}(t)}{\partial W_{\mathrm{att}}(t)}$ \\
        9:\hspace{0.9cm}  $W_{\mathrm{cla}}(t+1)$$\gets$$W_{\mathrm{cla}}(t)-\mu * \frac{\partial L_{A}(t)}{\partial W_{\mathrm{cla}}(t)}$ \\
        10:\hspace{0.6cm} end for\\
        11:\hspace{0.3cm} end for\\
        12: return $W_{t e m}$ ,$W_{s p a}$ ,$W_{att}$ ,$W_{class}$  \\
        \hline
        
    \end{tabular}
\end{table}

In the forward propagation process of network training, the input information is passed through the input layer and the hidden layer, processed layer by layer, and passed to the output layer. Our deep-and-shallow loss has three outputs, and the three cross-entropy losses of the three outputs and the label are calculated as the loss function. In step 4, $L_{T}$, $L_{S}$, $L_{A}$ represent temporal feature extraction network module loss, spatial feature extraction network module loss, and back-end classification module loss, respectively. We call $L_{T}$ and $L_{S}$ shallow loss, and $L_{A}$ stands for deep loss.  $L(t)$  represents a total loss. $\lambda_{1}$, $\lambda_{2}$ and $\left(1-\lambda_{1}-\lambda_{2}\right)$  respectively represent the weight assigned to each loss. The deep-and-shallow loss we designed has several advantages, first of all, for the fusion method of temporal and spatial features, deep-and-shallow loss assigns different weights to different network branches so that the network can focus more on difficult-to-train branches. Second, the spatial and temporal feature extraction network is optimized by the deep-and-shallow loss, which can speed up the convergence of the shallow network parameters to a certain extent.
 
 Then transfer to the back-propagation stage. In step 5, the three losses respectively calculate the partial derivative of the objective function to the weight of each neuron layer by layer, which constitutes the gradient of the objective function to the weight vector, which serves as the basis for modifying the weight. In steps 6-9, the training of the network is completed in the parameter update. $W_{t e m}$  represents the weight of the temporal feature extraction network. $W_{s p a}$ represents the weight of the spatial feature extraction network. $W_{att}$ represents the weight of the attention module, and $W_{cla}$ represents the weight of the back-end classification module. Our deep-and-shallow loss includes three losses. Therefore, it can be seen from steps 6-7 in Table 1 that the network needs to refer to the two output weights to modify the gradient when updating the network based on spatial information extraction and temporal information extraction.

\section*{4 Parallel spatial and temporal feature extraction network}
As shown in Fig. 2, the PSTNN (Parallel Spatial-Temporal Network) model is divided into two parts. One is a spatial information extractor described in section 4.1, and the other is a temporal information extractor described in section 4.2. Spatial information extractors include DNN, CNN, and ResNet designed by us. We designed LSTM and Bi-LSTM for temporal information extractors. The rest of this section will describe the network we designed in detail.

\begin{figure}[!htbp]
\includegraphics[width=4.5in]{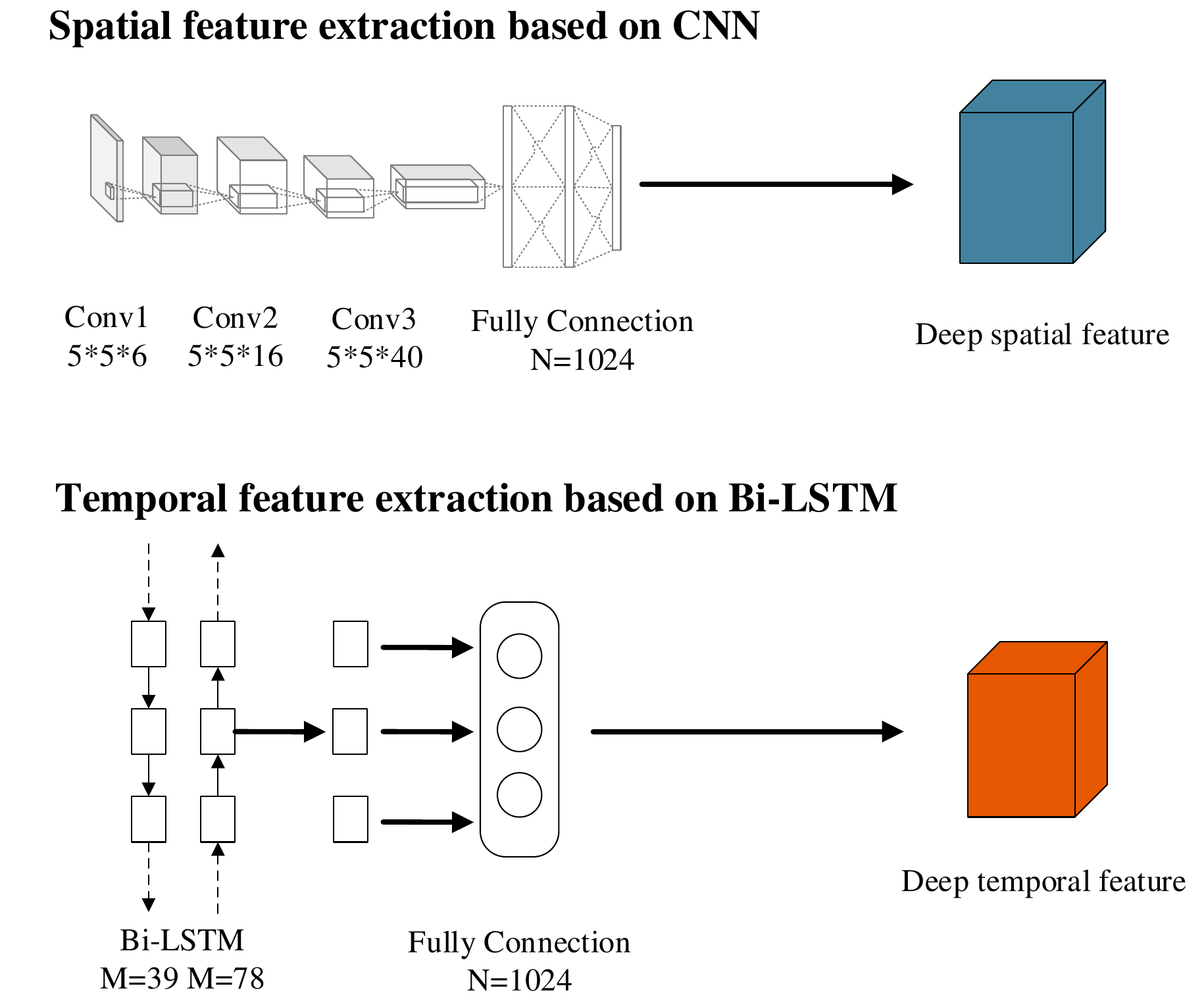}
  \caption{PSTNN network structure diagram, including spatial feature extraction network and temporal feature extraction network, where M represents the number of nodes of LSTM, and N represents the number of nodes of DNN.}
\end{figure}

\subsection*{4.1 Spatial feature extraction network}
We used the spatial feature extraction network, including DNN, CNN, and ResNet, based on the GSV feature. GSV is a C * F dimensional super vector connected by the GMM mean value. Assuming that the data of each category conforms to the Gaussian distribution, GMM is a spatial combination of multiple Gaussian distribution functions. GSV features can best represent the differences between various GMM so that GSV features can be regarded as spatial features. In section 4.1.1, we describe a DNN-based device source spatial information extraction network. In section 4.1.2, we design a CNN-based device source spatial information extraction network. In section 4.1.3, we design a residual network for the extraction of device source spatial information.

\subsubsection*{4.1.1 Spatial information feature extraction based on DNN}

DNN is usually used as a spatial feature extraction tool. In the DNN-based spatial feature extraction network, the input layer is a 2496-dimensional GSV feature. The design of multiple hidden layers is mainly to enhance the expressive ability of the model. Of course, the complexity of the multilayer perceptron model will increase. The output layer is the deep representation bottleneck feature based on DNN. To increase the nonlinear fitting ability of the model, we use the RELU activation function. The specific parameters of the DNN-based device source spatial feature extraction network we designed will be explained in section 6.3.


\subsubsection*{4.1.2 Spatial information feature extraction based on CNN}

In this section, we design a CNN to extract device source spatial information. Compared with DNN, the CNN performs better on high-dimensional data \cite{36}. CNN has fewer parameters, which shortens training time and improves operational efficiency. Its layer properties are more suitable for high-dimensional data. The CNN network structure mainly includes the convolutional layer, pooling layer, and batch normalization layer. The specific parameters of the CNN-based device source spatial feature extraction network we designed in section 6.3.

\subsubsection*{4.1.3 Spatial information featrue extraction based on ResNet}

In this section, we plan to deepen the network to obtain more prosperous device source spatial information. However, after increasing the number of network layers, the optimization effect worsens. In order to alleviate the problem of gradient explosion and gradient disappearance caused by network deepening \cite{33}, we design ResNet to extract the spatial information of the device source, the structure of which is shown in Fig. 3. Our residual network consists of four residual blocks. The input and output of each residual block will be connected to form a jump connection. Each residual block contains five layers of networks: four layers of convolution, one layer of pooling, one layer of batch normalization to reduce the risk of overfitting. See section 6.3 for specific network parameters.

\begin{figure}[!htbp]
\includegraphics[width=4.8in]{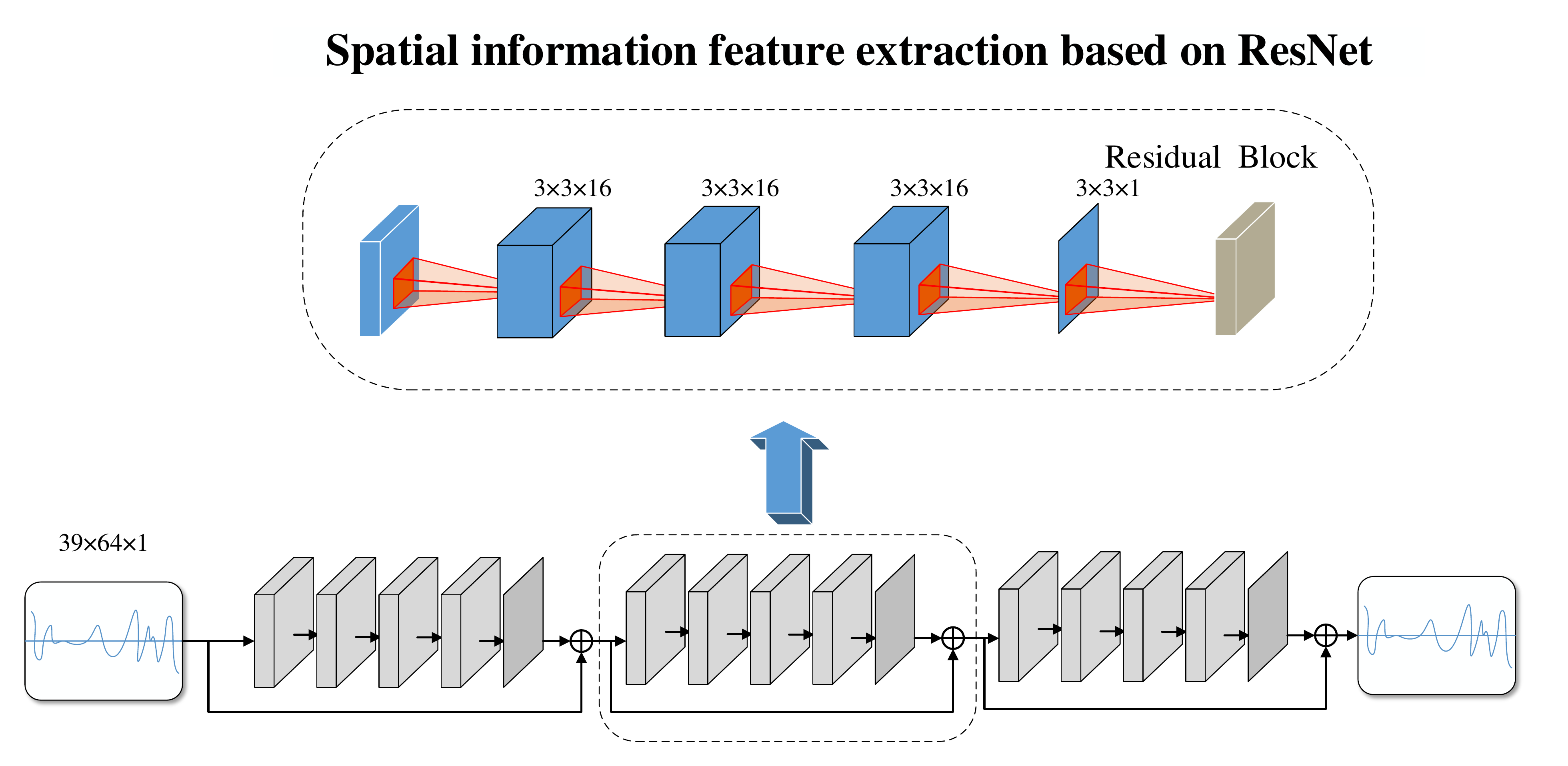}
  \caption{The figure shows the schematic diagram of the residual network and residual block parameters and structure.}
\end{figure}

\subsection*{4.2 Temporal feature extraction network}
In order to extract the temporal information from the device source, we designed LSTM and Bi-LSTM networks to extract the device source deep temporal features from MFCC. MFCC is a set of temporal feature vectors obtained by encoding the speech's physical information (spectral envelope and details). In section 4.2.1, we describe the LSTM based device source temporal information extraction network. In section 4.2.2, we introduce the Bi-LSTM based
device source temporal information extraction network.
\subsubsection*{4.2.1 Temporal information feature extraction based on LSTM}
In order to extract the temporal information from the device source, we design an LSTM + DNN network. The LSTM network refers to \cite{34}, which can solve the long-term dependence of temporal information and improve the adaptability of the network. The LSTM network can execute the temporal state of each unit through the gate structure, delete or add information, and essentially transmit useful device source information to the next frame. The structure diagram of LSTM + DNN we designed is shown in the following Fig. 4. The model takes the frame-level speech feature MFCC as input and obtains the output, corresponding to each frame through the 2-layer LSTM. M is the number of  LSTM hidden units, and N is the number of DNN hidden units.

\begin{figure}[!htbp]
\includegraphics[width=4.5in]{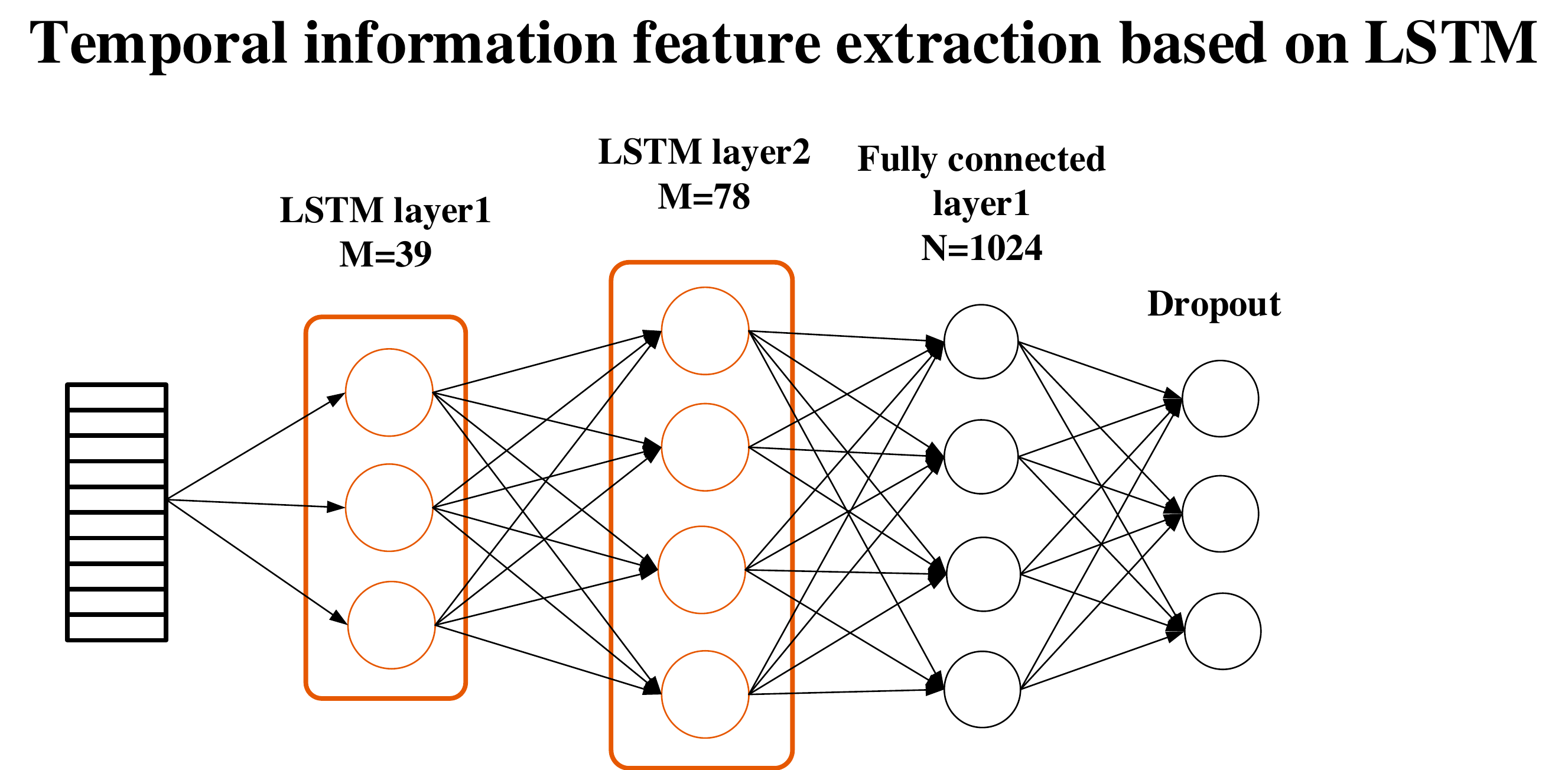}
  \caption{The figure is a schematic diagram of LSTM+DNN parameters and structure, where M represents the number of nodes of LSTM, and N represents the number of nodes of DNN.}
\end{figure}
\subsubsection*{4.2.2 Temporal information feature extraction based on Bi-LSTM}

Similarly, we also designed the Bi-LSTM temporal feature extractor. Bi-LSTM is the abbreviation of bidirectional long and short-term memory and combines forward LSTM and backward LSTM. Compared with the structure of LSTM + DNN, the prediction of Bi-LSTM + DNN is determined by the first inputs and the following inputs, which will be more accurate. However, it is undeniable that compared to LSTM, Bi-LSTM will be more challenging to converge and increase calculation.

\section*{5 Attention mechanism and back-end classification network}
This section describes the fusion mechanism for device source spatial information and temporal information and back-end classification network. The task of device source identification based on spatial information and temporal information has been completed, but the task based on the fusion of the two has not been developed. Considering the interdependence between spatial and temporal information in device source identification, in section 5.1, we use the attention mechanism to fuse two kinds of information. In section 5.2, we use a reduced DNN network for back-end classification network.

\subsection*{5.1 Attention mechanism}
Usually, using a single feature representation of the data may not be sufficient to complete the classification task. In this paper, we use two feature representations to capture different aspects of the input. The attention mechanism can assign significant weights to two representations, determining the most relevant aspects while ignoring noise and redundancy in the input. The expression of the attention mechanism is a weighted combination of two features and the weight of the attention mechanism. One advantage of the attention mechanism is to evaluate which combination of feature representations is the preferred expression for a specific classification task by checking the weights.

Essentially, the attention mechanism is a resource allocation mechanism that allocates available resources to more important features. The attention mechanism in this paper uses convolution, pooling, and activation functions (similar to softmax) to construct weights to readjust the feature map. Firstly, since the device source information of spatial is usually low-frequency, convolutional neural and average pooling operations can reduce the difference between adjacent GSVs and smooth features. Secondly, by adding a convolutional neural network, the network can introduce appropriate parameters to learn the best weights.

\begin{figure}[!htbp]
\includegraphics[width=4.5in]{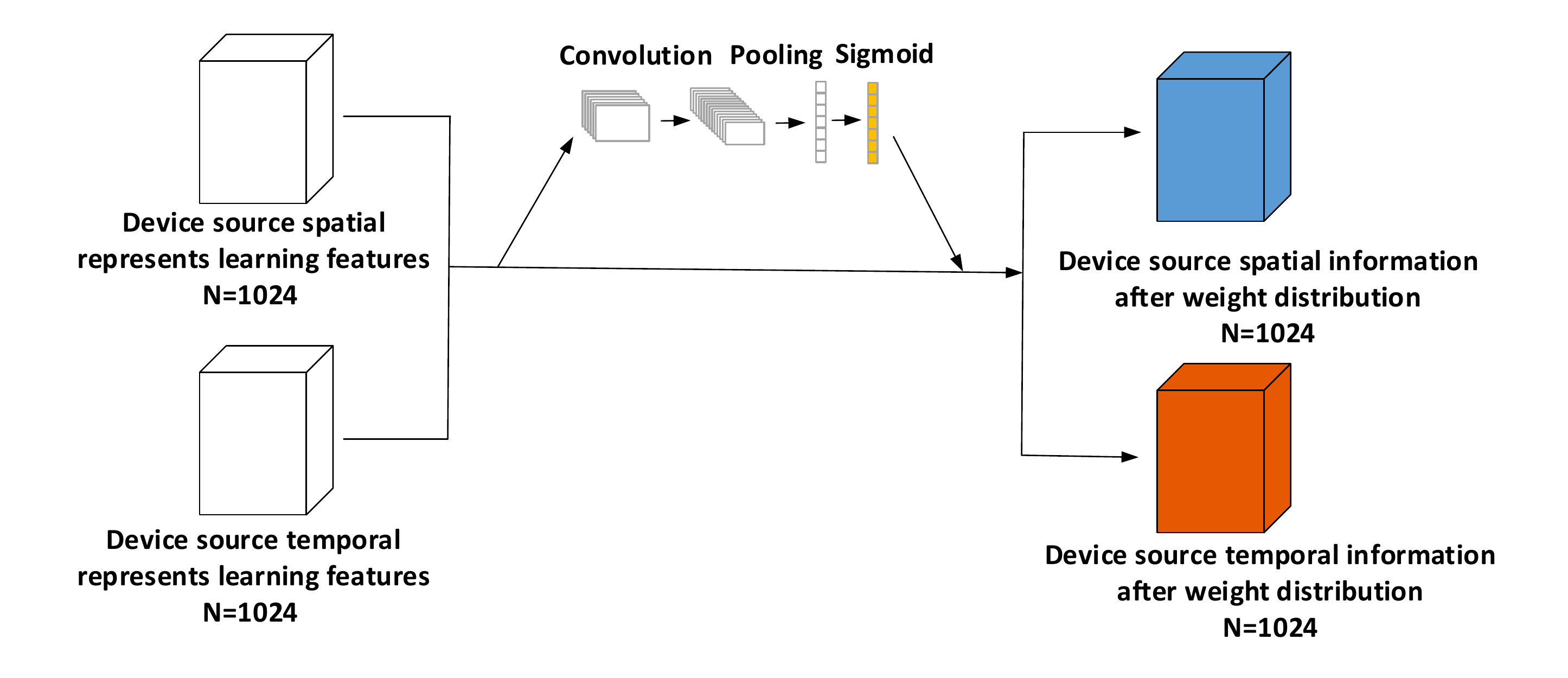}
  \caption{Schematic diagram of attention mechanism. N represents the dimension of the vector.}
\end{figure}

As shown in Figure 5, deep device source spatial and temporal features are both 1024-dimensional vectors. They will be merged into a 1024*2 matrix for the convolutional layer and the pooling layer to build weights before the attention mechanism. In the process of the attention mechanism, the network learns parameters and generates weight values through the convolutional layer and the pooling layer, and the weight values are multiplied by the temporal feature map and the spatial feature map and merged to obtain the final feature. After the attention mechanism, we obtain the fusion vector feature of the spatial and temporal representation of the device source, and their dimensions are 1024-dimensional vectors. The fusion feature discards redundant information and magnifies the inter-class representation capabilities of the two representation features.

\subsection*{5.2 Back-end classification network}
In the back-end classification network, we use a DNN network with decreasing neurons. Mainly reduce the storage and calculation cost of parameters or calculation results in the model and increase the fit. The goal is to map feature data to a series of quantitative trim levels. The DNN network with decreasing neurons can effectively reduce parameter redundancy and help deploy deep learning applications. The parameter designed according to our previous work \cite{13}.

\section*{6 Experimental results and discussion}
In this section, we introduce the dataset, baseline system, experimental setup, and experimental results. In order to verify the effectiveness of our work, we design six sets of experiments to verify our contribution: (1)Comparison experiment based on UBM Gaussian number. (2)Comparison experiment based on back-end classification. (3)Network selection comparison experiment based on PSTNN. (4) Comparison experiment based on our proposed deep-and-shallow loss weight setting and single loss function. (5) comparison experiment based on the effectiveness of the attention mechanism. (6) Compare the end-to-end model with the GMM-UBM \cite{5}, GSV-SVM \cite{37}, BED-SVM \cite{28}, and previous work \cite{13}.

\subsection*{6.1 Dataset description}
To verify the performance of the proposed model, this paper uses the CCNU-Mobile dataset as the experimental dataset. We established the CCNU-Mobile dataset in 2019. The CCNU-Mobile dataset contains 8 different brands and 45 different devices. See Table 2 for specific models and
brands. In addition, the CCNU-Mobile dataset is recorded in the TIMIT dataset, each mobile device has 642 recording files, and the sample size of the dataset is 45*642 = 28890. To ensure the recording quality, the recording environment is a professional recording studio that is almost completely quiet. The recorded voice data sampling rate is 32Khz, and the quantization rate is 16bit.

\begin{table}[!htbp]
\caption{Brands and models of the CCNU-Mobile dataset
.}
\centering
    \begin{tabular}{c c}
        \hline
        \tabincell{c}{{Brands}} & \tabincell{c}{
        Models} \\      
        \hline
        \tabincell{c}{APPLE}  & \tabincell{c}{{iphone6(4),iphone6s(3),iphoneSE, ipad7,iphone7p,iphoneX,air2(2),air1}}  \\
        \hline
        \tabincell{c}{HUAWEI}  & \tabincell{c}{{tagal00, nova,novo2s,nova3e,honor7x,honor8(3),honorV8, honor9,honor10,p10,p20}}  \\
        \hline
        
        \tabincell{c}{XIAOMI}  & \tabincell{c}{{mi2s, note3, mi5,mi8,mi8se(2),mix2,redmiNote4x,redmi3S}}  \\
        \hline
         \tabincell{c}{VIVO}  & \tabincell{c}{{y11t,x3f,x7}}  \\
         \hline
         \tabincell{c}{ZTE}  & \tabincell{c}{{c880a,g719c}}  \\
         \hline
         \tabincell{c}{SAMSUNG}  & \tabincell{c}{{sphd710,s8}}  \\
        \hline
         \tabincell{c}{OPPO}  & \tabincell{c}{r9s}  \\
         \hline
         \tabincell{c}{NUBIA}  & \tabincell{c}{z11}  \\
        \hline

    \end{tabular}
\end{table}

\subsection*{6.2 
Experimental Setup}
This work divides the dataset into 8:2, each training device contains 514 recordings, and the test device contains 128 recordings. We use a batch normalization layer in the network, which speeds up the convergence speed of the model and, more importantly, alleviates the overfitting problem in the deep network to a certain extent. We have added Dropout to the network. Dropout can effectively alleviate the occurrence of over-fitting and achieve the effect of regularization to a certain extent. We use Adam's optimization method to optimize all network parameters, and the exponential decay rates of the first and second moment are 0.9 and 0.999. We use Tensorflow and Keras software packages to train and test all network models in this work. The GPU hardware information is RTX TITAN.
When extracting MFCC in the experiment, the length of the signal frame is 30 ms, and the overlap time is 15 ms. In order to make the feature better reflect the time domain continuity, we also added the first-order difference and the second-order difference in the MFCC. Therefore we use 39-dimensional MFCC and TIMIT dataset to train 64, 128, 256-dimensional UBM. GSV uses 64, 128, 256 Gaussian 39-dimensional parameters, and the baseline system uses GMM-UBM \cite{5}, GSV-SVM \cite{37}, and BED-SVM models \cite{28}. All the experimental evaluation indicators in this paper use classification accuracy (ACC), and the total number of categories is 45.

\subsection*{6.3 Hyperparameter settings of the network}
 This section mainly introduces the parameter composition of the PSTNN network. As shown in Table 3, the SFENN (spatial feature extraction network) we designed includes DNN, CNN, and ResNet. The parameters of each network are shown in Table 3. The input dimension of DNN is 2496, and the three hidden layer nodes are 1024. The CNN network contains three layers of convolution, three layers of pooling, and one layer of full connection. The number of the three-layer convolution kernel is 6, 16, and 40, respectively, and the size of the convolution kernel is 5*5. All pooling layers in this paper adopt maximum pooling. The number of hidden layers of the fully connected layer is 1024. The residual network uses four residual blocks, and each residual block includes four convolutional layers, four Batch Normalization layers, and a pooling layer. The convolution kernel of the convolution layer is 16, and the size of the convolution kernel is 3*3. TFENN (temporal feature extraction network) includes LSTM and Bi-LSTM. LSTM includes two hidden layers and a fully connected layer. The input has the dimension of [128,64,39], where 128 is the batch size, 64 is the number of frames, and 39 is the number of features exacted from speech. The number of units in the first hidden layer is 39, and the number of units in the second hidden layer is 78. Bi-LSTM is similar to LSTM.

\begin{table*}[!htbp]
\centering
\caption{PSTNN structure design parameter.}

    \begin{tabular}{c c c c c}
        \hline
        
        
        \multicolumn{3}{c}{SFENN} &
        \multicolumn{2}{c}{TFENN}\\
        \tabincell{c}{DNN}  & \tabincell{c}{CNN} & \tabincell{c}{ResNet}& \tabincell{c}{LSTM}& \tabincell{c}{Bi-LSTM} \\
        \hline
        \tabincell{c}{{Input (2496)}}  & \tabincell{c}{Input (64*39*1)} & \tabincell{c}{Input}& \tabincell{c}{Input}& \tabincell{c}{Input} \\
        \hline
        \tabincell{c}{{FC (1024)}}  & \tabincell{c}{Conv (6*5*5)} & \tabincell{c}{BN+Relu}& \tabincell{c}{LSTM(39)}& \tabincell{c}{Bi-LSTM(39)} \\
        \hline
        \tabincell{c}{{FC (1024)}}  & \tabincell{c}{pooling } & \tabincell{c}{Conv(16*3*3)}& \tabincell{c}{LSTM(78)}& \tabincell{c}Bi-LSTM(78){} \\
        \hline
        \tabincell{c}{{FC (1024)}}  & \tabincell{c}{Conv (16*5*5)} & \tabincell{c}{BN+Relu}& \tabincell{c}{FC(1024)}& \tabincell{c}{FC(1024)}\\
        \hline
        \tabincell{c}{}  & \tabincell{c}{pooling } & \tabincell{c}{Conv(16*3*3)}& \tabincell{c}{}& \tabincell{c}{}\\
        \hline
        \tabincell{c}{}  & \tabincell{c}{Conv (40*5*5)} & \tabincell{c}{BN+Relu}& \tabincell{c}{}& \tabincell{c}{}\\
        \hline
        \tabincell{c}{}  & \tabincell{c}{pooling } & \tabincell{c}{Conv(16*3*3)}& \tabincell{c}{}& \tabincell{c}{}\\
        \hline
        \tabincell{c}{}  & \tabincell{c}{FC (1024)} & \tabincell{c}{BN+Relu}& \tabincell{c}{}& \tabincell{c}{}\\
        \hline
        \tabincell{c}{}  & \tabincell{c}{} & \tabincell{c}{Conv(16*3*3)}& \tabincell{c}{}& \tabincell{c}{}\\
        \hline
        \tabincell{c}{}  & \tabincell{c}{} & \tabincell{c}{pooling (Input)}& \tabincell{c}{}& \tabincell{c}{}\\
        \hline
        \tabincell{c}{}  & \tabincell{c}{} & \tabincell{c}{Add()}& \tabincell{c}{}& \tabincell{c}{}\\
        \hline
        
    \end{tabular}
\end{table*}

\subsection*{6.4 Comparison experiment based on UBM Gaussian number}
To explore the impact of UBM on the model's overall performance, first, we used the TIMIT dataset to extract 64, 128, and 256 Gaussian UBM. Then, using the MAP algorithm to extract the GSV features of each recording sample, the MFCC is consistent with section 6.2. The PSTNN network is a combination of DNN + Bi-LSTM. Finally, the experimental results are shown in Table 4.

\begin{table}[!htbp]
\caption{Comparison experiment based on the effectiveness of UBM Gaussian number.}
\centering
    \begin{tabular}{c c c}
        
        \hline
        \tabincell{c}{{UBM Gaussian number}} &\tabincell{c}{network parameter} & \tabincell{c}{{DNN + Bi-LSTM} 
        \\{(ACC)}} \\      
        \hline
        \tabincell{c}{{64}}  & \tabincell{c}{114765k} & \tabincell{c}{97.5\%} \\
        \hline
        \tabincell{c}{{128}}  & \tabincell{c}{120995k} & \tabincell{c}{97.3\%} \\
        \hline
        \tabincell{c}{{256}}  & \tabincell{c}{133455k} & \tabincell{c}{97.0\%} \\
        \hline
        
    \end{tabular}
\end{table}

It can be seen from the experimental results that when the Gaussian number is 64, the model has the best effect, reaching 97.5\%, and the network parameter amount is 114765k. It is proved that the increase of the Gaussian number will increase the amount of calculation and cause redundancy to the data. The main reason is that the category of the data set is 45 categories, and 64 Gauss is enough to characterize the similarities and differences of the 45 categories of data.

\subsection*{6.5 Comparison experiment based on back-end classification}
To verify the impact of the back-end classification network on the model performance, we conducted a comparison experiment of the back-end classification network. The parameters of the back-end classification network are simple softmax classification, 1-layer full connection, 2-layer full connection, and 3-layer full connection.

\begin{table}[!htbp]
\caption{Comparison experiment based on back-end classification.}
\centering
    \begin{tabular}{c c}
        
        \hline
        \tabincell{c}{{back-end classification}}  & \tabincell{c}{{DNN + Bi-LSTM}} 
        \\{(ACC)} \\      
        \hline
        \tabincell{c}{{simple softmax classification}}  &  \tabincell{c}{97.4\%} \\
        \hline
        \tabincell{c}{{1-layer full connection}}  &  \tabincell{c}{97.4\%} \\
        \hline
        \tabincell{c}{{2-layer full connection}}  &  \tabincell{c}{97.5\%} \\
        \hline
         \tabincell{c}{{3-layer full connection}}  &  \tabincell{c}{97.5\%} \\
        \hline
    \end{tabular}
\end{table}

The experimental results in Table 5 show that when the back-end classification networks are 2-layer and 3-layer fully connection, the model has the best effect, which is 97.5\%. It is 0.1\% higher than the simple softmax network, so when the back-end classification is 2-layer fully connection, the efficiency is the highest, and it proves that adding an appropriate number of fully connected layers can increase the complexity of the model and improve the learning ability of the network model.

\subsection*{6.6 Network selection comparison experiment based on PSTNN}
To optimize our network structure, we design an end-to-end framework comparison experiment based on PSTNN, which compares the networks that extract spatial and temporal features. The weight ratio of deep-and-shallow loss is 0.25:0.5:0.25.
\begin{table*}[!htbp]
\centering
\caption{Comparison and optimization experiment on network selection in PSTNN.}
    \begin{tabular}{c c c c c c c c}
        \hline
        
        \multirow{2}{*}{End-to-end} &
        \multicolumn{3}{c}{SFENN} &
        \multicolumn{2}{c}{{TFENN}} & \multicolumn{2}{c}{Epoch}\\
          & DNN & CNN & ResNet & LSTM & Bi-LSTM & 100ep &200ep \\
        \hline
        \tabincell{c}{{Experiment 1}}  & \tabincell{c}{\Checkmark} & \tabincell{c}{}& \tabincell{c}{}& \tabincell{c}{}& \tabincell{c}{}& \tabincell{c}{96.5\%}& \tabincell{c}{96.6\%}\\
        \hline
        \tabincell{c}{{Experiment 2}}  & \tabincell{c}{} & \tabincell{c}{\Checkmark}& \tabincell{c}{}& \tabincell{c}{}& \tabincell{c}{}& \tabincell{c}{94.5\%}& \tabincell{c}{94.6\%}\\
        \hline
        \tabincell{c}{{Experiment 3}}  & \tabincell{c}{} & \tabincell{c}{}& \tabincell{c}{\Checkmark}& \tabincell{c}{}& \tabincell{c}{}& \tabincell{c}{94.6\%}& \tabincell{c}{94.5\%}\\
        \hline
        \tabincell{c}{{Experiment 4}}  & \tabincell{c}{} & \tabincell{c}{}& \tabincell{c}{}& \tabincell{c}{\Checkmark}& \tabincell{c}{}& \tabincell{c}{68.8\%}& \tabincell{c}{69.4\%}\\
        \hline
        \tabincell{c}{{Experiment 5}}  & \tabincell{c}{} & \tabincell{c}{}& \tabincell{c}{}& \tabincell{c}{}& \tabincell{c}{\Checkmark}& \tabincell{c}{70.9\%}& \tabincell{c}{70.7\%}\\
        \hline
        \tabincell{c}{{Experiment 6}}  & \tabincell{c}{\Checkmark} & \tabincell{c}{}& \tabincell{c}{}& \tabincell{c}{\Checkmark}& \tabincell{c}{}& \tabincell{c}{97.3\%}& \tabincell{c}{97.3\%}\\
        \hline
        \tabincell{c}{{Experiment 7}}  & \tabincell{c}{} & \tabincell{c}{\Checkmark}& \tabincell{c}{}& \tabincell{c}{\Checkmark}& \tabincell{c}{}& \tabincell{c}{96.2\%}& \tabincell{c}{96.1\%}\\
        \hline
        \tabincell{c}{{Experiment 8}}  & \tabincell{c}{} & \tabincell{c}{}& \tabincell{c}{\Checkmark}& \tabincell{c}{\Checkmark}& \tabincell{c}{}& \tabincell{c}{96.2\%}& \tabincell{c}{96.2\%}\\
        \hline
        \tabincell{c}{{Experiment 9}}  & \tabincell{c}{\Checkmark} & \tabincell{c}{}& \tabincell{c}{}& \tabincell{c}{}& \tabincell{c}{\Checkmark}& \tabincell{c}{97.2\%}& \tabincell{c}{97.5\%}\\
        \hline
        \tabincell{c}{{Experiment 10}}  & \tabincell{c}{} & \tabincell{c}{\Checkmark}& \tabincell{c}{}& \tabincell{c}{}& \tabincell{c}{\Checkmark}& \tabincell{c}{96.1\%}& \tabincell{c}{96.3\%}\\
        \hline
        \tabincell{c}{{Experiment 11}}  & \tabincell{c}{} & \tabincell{c}{}& \tabincell{c}{\Checkmark}& \tabincell{c}{}& \tabincell{c}{\Checkmark}& \tabincell{c}{96.0\%}& \tabincell{c}{96.2\%}\\
        \hline
        
    \end{tabular}
\end{table*}

The comparison results of the PSTNN and the single feature extraction network are shown in Table 6. According to the longitudinal comparison in Table 6, when the training period is 100 epochs, DNN is best used as a spatial extraction network, followed by ResNet, and finally, the CNN model. In the temporal network extractor, LSTM is slightly better than Bi-LSTM. When the training period is 200 epochs, this situation is changed, and the effect of using Bi-LSTM exceeds the effect of LSTM. It proves that Bi-LSTM needs longer training time than LSTM network to achieve convergence and balance. From a horizontal comparison, DNN is used to extract the device source spatial information for the single network model to achieve the best result of 96.6\%. The DNN + Bi-LSTM network combination achieves the best results with an accuracy rate of 97.5\% for the fusion network model. Compared with another fused network model, the maximum increase is 1.4\%. From the comparison of Experiment 3 and Experiment 11, the ResNet + Bi-LSTM combination is better than the single ResNet spatial feature extraction network, which proves that the fusion network is better than one of the single feature extraction networks. However, from Experiment 1 and Experiment 11, the effect of a single DNN spatial feature is better than the ResNet + Bi-LSTM combination, which proves that each spatial feature extraction network has different spatial extraction capabilities. At the same time, it also proves that DNN spatial feature extraction ability is much better than ResNet so that the combination of ResNet + Bi-LSTM is also lower than DNN spatial feature extraction effect. All in all, our DNN + Bi-LSTM combination achieves the best fitting effect, which is 27\% higher than the single network model. The reason for this problem is that it combines the spatial and temporal information in the device source and further proves that our method is effective.

\subsection*{6.7 Comparison experiment based on our proposed deep-and-shallow loss} weight setting and single loss function
To explore the effectiveness of our proposed deep-and-shallow loss by controlling the weight of deep-and-shallow loss, we design several sets of comparative experiments to optimize our model. In addition, to verify the effectiveness of the deep-and-shallow loss, we compare the deep-and-shallow loss with the use of a single cross-entropy loss function. The PSTNN network uses DNN + Bi-LSTM and DNN-LSTM, which have better performance in Section 6.6 experiment, and is trained for 200 epochs. The learning rate benchmark is set to 0.001 and is reduced to 1/10 every 20 cycles. 
 
\begin{table}[!htbp]
\caption{Comparison experiment of weight selection based on deep-and-shallow loss and single loss.}
\centering
    \begin{tabular}{c c c}
        \hline
        \tabincell{c}{{Loss function and parameters}} & \tabincell{c}{{Network model}} & \tabincell{c}{{End-to-end}\\{(ACC)}} \\      
        \hline
        \tabincell{c}{{Categorical crossentropy loss}}  & \tabincell{c}{{DNN-LSTM} \\{DNN + Bi-LSTM}} & \tabincell{c}{{96.5\%} \\{96.6\%}}\\
        \hline
        \tabincell{c}{{deep-and-shallow loss} (0.25:0.5:0.25)}  & \tabincell{c}{{DNN-LSTM} \\{DNN + Bi-LSTM}} & \tabincell{c}{{97.3\%} \\{97.5\%}}\\
        \hline
        \tabincell{c}{{deep-and-shallow loss} (0.4:0.2:0.4)}  & \tabincell{c}{{DNN-LSTM} \\{DNN + Bi-LSTM}} & \tabincell{c}{{97.2\%} \\{97.2\%}}\\
        \hline
        \tabincell{c}{{deep-and-shallow loss} (0.25:0.25:0.5)}  & \tabincell{c}{{DNN-LSTM} \\{DNN + Bi-LSTM}} & \tabincell{c}{{97.1\%} \\{97.5\%}}\\
        \hline
        \tabincell{c}{{deep-and-shallow loss} (0.2:0.6:0.2)} & \tabincell{c}{{DNN-LSTM} \\{DNN + Bi-LSTM}} & \tabincell{c}{{97.3\%} \\{97.3\%}}\\
        \hline
    \end{tabular}
\end{table}

Table 7 shows the experimental results of end-to-end classification based on deep-and-shallow loss weight distribution. When using deep-and-shallow loss to optimize the network, all the results are higher than 97.1\% regardless of the weight of loss. Compared with the network model using a single loss, our proposed deep-and-shallow loss is significantly better than the single loss network model, and the maximum increase accuracy is 1\%. It shows that by co-optimizing the shallow loss and deep loss, the network can converge better, and the fit of the network can be increased. By setting up a comparative experiment with different weights of deep-and-shallow loss, we found the optimal weight distribution of our proposed PSTNN. When the weight distribution is 0.25:0.5:0.25 and 0.25:0.25:0.5, our network model achieves the best result, reaching 97.5\%. The results show that better results can be obtained when appropriate weights are assigned to the network. It may be because Bi-LSTM is more difficult to train and adapt than DNN and CNN networks, so it needs to directly or indirectly increase the training loss weight.

\subsection*{6.8 The attention mechanism verifies the effectiveness of the combination of temporal and spatial features}
To verify the influence of the attention mechanism on our PSTNN network. We design a comparison experiment between the splicing feature and feature using the attention mechanism. The splicing feature network removes the attention module based on the PSTNN, splice the SFENN and TFENN features before and after. The experimental parameters are consistent with those in section  6.3.

\begin{table}[!htbp]
\caption{Comparative experiment based on the effectiveness of the attention mechanism.}
\centering
    \begin{tabular}{c c}
        
        \hline
        \tabincell{c}{{Network model}} & \tabincell{c}{{End-to-end}
        \\{(ACC)}} \\      
        \hline
        \tabincell{c}{{DNN + Bi-LSTM + splicing}}  & \tabincell{c}{97.3\%}  \\
        \hline
        \tabincell{c}{{DNN + Bi-LSTM + attention}}  & \tabincell{c}{97.5\%}  \\
        \hline
        
    \end{tabular}
\end{table}
It can be seen from Table 8 that 0.2\% improves the accuracy of the attention mechanism compared with the accuracy of the simple splicing network structure. It proves that the attention mechanism is effective for different types of feature weight distribution. On the one hand, the convolutional layer in the attention mechanism provides additional learning parameters, thereby improving the fitting ability of the network. On the other hand, the attention mechanism achieves feature selection, which provides greater weight for essential features and leads to better results.
\subsection*{6.9 Compare the end-to-end model with the baseline and previous work}
To verify that the end-to-end model we proposed is better than the traditional model and previous work \cite{13}, we design an experiment to compare the end-to-end network with BED-SVM, GMM-UBM, GSV-SVM, and previous work. The PSTNN in the end-to-end network contains a combination of DNN + Bi-LSTM, and the parameters are the same as in section 6.3. The BED+SVM model is the same as the parameters in \cite{28}.

\begin{table}[!htbp]
\caption{Verification of effectiveness based on end-to-end structure.}
\centering
    \begin{tabular}{c c c c}
        
        \hline
        \tabincell{c}{{Model}} &
        \tabincell{c}{Features} &
        \tabincell{c}{{Classification results}\\{(ACC)}} & \tabincell{c}{{Testing time}\\{(second)}} \\
        
        \hline
        \tabincell{c}{{GMM-UBM}}  &
        \tabincell{c}{MFCC}&
        \tabincell{c}{31.5\%}
        &
        \tabincell{c}{0.0677}
        \\
        \hline
        \tabincell{c}{{GSV-SVM}}  &
        \tabincell{c}{GSV} &
        \tabincell{c}{93.6\%}
        &
        \tabincell{c}{0.0519}\\
        \hline
        \tabincell{c}{{BED-SVM}}  & 
        \tabincell{c}{BED} &
        \tabincell{c}{92.5\%}
        &
        \tabincell{c}{0.0026}
        \\
        \hline
        \tabincell{c}{{previous work}}  &
        \tabincell{c}{MFCC,GSV,I-vector} 
        &
        \tabincell{c}{97.3\%} 
        &
        \tabincell{c}{0.0017} 
        \\
        \hline
         \tabincell{c}{{this work}}  &
         \tabincell{c}{MFCC,GSV} &
         \tabincell{c}{97.5\%} 
         &
        \tabincell{c}{0.0213}
         \\
        \hline
        
    \end{tabular}
\end{table}

It can be seen from Table 9. The results show that this work is 66 \% higher than the GMM-UBM model and is 4 \% higher than the GSV-SVM model, which is 5 \% higher than the BED-SVM model. It proves that our end-to-end framework performs better than the traditional method, which shows that our end-to-end framework can better play the performance of the entire model compared to the traditional model by jointly optimizing model weights. In addition, The number of features of the work in this paper is lower than the previous work, but the accuracy is indeed improved, which more intuitively proves our contribution compared to the previous work.

In terms of test time complexity, the index of this paper adopts the time of testing a single sample. As can be seen from Table 9, the test time complexity of this paper is ahead of the GMM-UBM model and GSV-SVM model, which is faster than the GMM-UBM model test time. It is 0.046 seconds faster, which is 0.03 seconds faster than the GSV-SVM model. The test time complexity of this work is 0.02 seconds behind the previous work and 0.019 seconds behind the BED-SVM model. It proves that the work of this paper achieves a better accuracy improvement when the time complexity gap is not significant.
\subsection*{6.10 Discussion}

In this section, we design six sets of experiments based on our four contributions. The first set of experiments verified the influence of the number of UBM on the entire model. The experiment proved that the 64-dimensional UBM had the best effect. The second set of experiments discussed the impact of the back-end classification network on the model. The experiment proved that the two-layer full connection is the most efficient. The third set of experiments mainly discusses spatial and temporal feature extraction networks and compares them with a single network, and the experiment proves that our proposed combination of spatial and temporal features achieves the best results. The fourth group of experiments mainly conduct experiments on the optimal weight distribution of deep-and-shallow loss and compare it with a single loss. The experimental results show that our deep-and-shallow loss has better performance than a single loss. The fifth set of experiments compares the attention mechanism's effect, which proves that the attention mechanism achieves the best results. The sixth set of experiments verifies the effectiveness of the end-to-end model from an overall perspective by comparing it with traditional methods and previous work.

\section*{7 Conclusions}
This paper proposes a device source identification method based on an end-to-end framework, which combines spatial and temporal information. first, we propose a fusion system of spatial and temporal features, which combines the spatial and temporal features in device source identification. Secondly, we proposed a deep-and-shallow loss and discussed the optimal weight of these loss combinations. Third, we add the attention mechanism to the fusion of spatial and temporal information. Fourth, we have established an end-to-end device source identification task, using deep-and-shallow optimization to make the system more compact and mobile. In summary, after our experimental exploration, the final solution to the problem in this article is the combination of DNN + Bi-LSTM for recording device source identification.

Although the work in this paper has achieved good results, the computational cost is relatively high. On the one hand, due to LSTM and Bi-LSTM networks in the temporal feature extraction module, they perform well in processing temporal data. However, their parameters are extensive, resulting in too long training time. On the other hand, the structure of the model is also more complicated. Therefore, future work will study how to improve the accuracy of device source identification while reducing computational costs.


\begin{backmatter}
	
\section*{Abbreviations}
CNN: Convolutional Neural Networks; DNN: Deep Neural Networks; LSTM: Long Short-Term Memory; SVM: Support Vector Machine; MFCC: Mel Frequency Cepstral Coefficient; BFCC: Bark Frequency Cepstrum Coefficient; LPCC: Linear Prediction Cepstrum Coefficient; PNCC: Power Normalized Cepstrum Coefficient; GSV: Gaussian Super Vector; GMM: Gaussian Mixture Model; SRC: Sparse Representation-based Classifier; RBF: Radial Basis Function kernel; GLDS: Generalized Linear Discriminant Sequence kernel; PSTNN: parallel spatial-temporal network

\section*{Funding}
This research was supported by National Natural Science Foundation of China (No.61901165, 61501199), Science and Technology Research Project of Hubei Education Department (No. Q20191406), Hubei Natural Science Foundation (No. 2017CFB683), and self-determined research funds of CCNU from the colleges basic research and operation of MOE (No. CCNU20ZT010).

\section*{Availability of data and materials}
Please contact authors for data requests.

\section*{Competing interests}
The authors declare that they have no competing interests.

\section*{Authors' contributions}
Equal contribution from all authors. All authors read and approved the final manuscript.

\section*{Acknowledgements}
The authors acknowledge the comments by anonymous reviewers that helped to improve a preliminary version of the paper.


\bibliographystyle{bmc-mathphys} 
\bibliography{bmc_article,MyWork}      


\begin{thebibliography}{56}
\ifx \bisbn   \undefined \def \bisbn  #1{ISBN #1}\fi
\ifx \binits  \undefined \def \binits#1{#1}\fi
\ifx \bauthor  \undefined \def \bauthor#1{#1}\fi
\ifx \batitle  \undefined \def \batitle#1{#1}\fi
\ifx \bjtitle  \undefined \def \bjtitle#1{#1}\fi
\ifx \bvolume  \undefined \def \bvolume#1{\textbf{#1}}\fi
\ifx \byear  \undefined \def \byear#1{#1}\fi
\ifx \bissue  \undefined \def \bissue#1{#1}\fi
\ifx \bfpage  \undefined \def \bfpage#1{#1}\fi
\ifx \blpage  \undefined \def \blpage #1{#1}\fi
\ifx \burl  \undefined \def \burl#1{\textsf{#1}}\fi
\ifx \doiurl  \undefined \def \doiurl#1{\textsf{#1}}\fi
\ifx \betal  \undefined \def \betal{\textit{et al.}}\fi
\ifx \binstitute  \undefined \def \binstitute#1{#1}\fi
\ifx \binstitutionaled  \undefined \def \binstitutionaled#1{#1}\fi
\ifx \bctitle  \undefined \def \bctitle#1{#1}\fi
\ifx \beditor  \undefined \def \beditor#1{#1}\fi
\ifx \bpublisher  \undefined \def \bpublisher#1{#1}\fi
\ifx \bbtitle  \undefined \def \bbtitle#1{#1}\fi
\ifx \bedition  \undefined \def \bedition#1{#1}\fi
\ifx \bseriesno  \undefined \def \bseriesno#1{#1}\fi
\ifx \blocation  \undefined \def \blocation#1{#1}\fi
\ifx \bsertitle  \undefined \def \bsertitle#1{#1}\fi
\ifx \bsnm \undefined \def \bsnm#1{#1}\fi
\ifx \bsuffix \undefined \def \bsuffix#1{#1}\fi
\ifx \bparticle \undefined \def \bparticle#1{#1}\fi
\ifx \barticle \undefined \def \barticle#1{#1}\fi
\ifx \bconfdate \undefined \def \bconfdate #1{#1}\fi
\ifx \botherref \undefined \def \botherref #1{#1}\fi
\ifx \url \undefined \def \url#1{\textsf{#1}}\fi
\ifx \bchapter \undefined \def \bchapter#1{#1}\fi
\ifx \bbook \undefined \def \bbook#1{#1}\fi
\ifx \bcomment \undefined \def \bcomment#1{#1}\fi
\ifx \oauthor \undefined \def \oauthor#1{#1}\fi
\ifx \citeauthoryear \undefined \def \citeauthoryear#1{#1}\fi
\ifx \endbibitem  \undefined \def \endbibitem {}\fi
\ifx \bconflocation  \undefined \def \bconflocation#1{#1}\fi
\ifx \arxivurl  \undefined \def \arxivurl#1{\textsf{#1}}\fi
\csname PreBibitemsHook\endcsname

\bibitem{Zeng2021a}
\begin{barticle}
\bauthor{\bsnm{Zeng}, \binits{C.}},
\bauthor{\bsnm{Zhu}, \binits{D.}},
\bauthor{\bsnm{Wang}, \binits{Z.}},
\bauthor{\bsnm{Wu}, \binits{M.}},
\bauthor{\bsnm{Xiong}, \binits{W.}},
\bauthor{\bsnm{Zhao}, \binits{N.}}:
\batitle{Spatial and temporal learning representation for end-to-end recording
  device identification}.
\bjtitle{EURASIP Journal on Advances in Signal Processing}
\bvolume{2021}(\bissue{1}),
\bfpage{41}
(\byear{2021}).
doi:\doiurl{10.1186/s13634-021-00763-1}
\end{barticle}
\endbibitem

\bibitem{Wang2022t}
\begin{barticle}
\bauthor{\bsnm{Wang}, \binits{Z.}},
\bauthor{\bsnm{Yang}, \binits{Y.}},
\bauthor{\bsnm{Zeng}, \binits{C.}},
\bauthor{\bsnm{Kong}, \binits{S.}},
\bauthor{\bsnm{Feng}, \binits{S.}},
\bauthor{\bsnm{Zhao}, \binits{N.}}:
\batitle{Shallow and deep feature fusion for digital audio tampering
  detection}.
\bjtitle{EURASIP Journal on Advances in Signal Processing}
\bvolume{2022}(\bissue{69}),
\bfpage{1}--\blpage{20}
(\byear{2022}).
doi:\doiurl{10.1186/s13634-022-00900-4}
\end{barticle}
\endbibitem

\bibitem{1}
\begin{bchapter}
\bauthor{\bsnm{Narkhede}, \binits{M.}},
\bauthor{\bsnm{Patole}, \binits{R.}}:
\bctitle{Acoustic scene identification for audio authentication}.
In: \beditor{\bsnm{Wang}, \binits{J.}},
\beditor{\bsnm{Reddy}, \binits{G.R.M.}},
\beditor{\bsnm{Prasad}, \binits{V.K.}},
\beditor{\bsnm{Reddy}, \binits{V.S.}} (eds.)
\bbtitle{Soft Computing and Signal Processing},
pp. \bfpage{593}--\blpage{602}
(\byear{2019})
\end{bchapter}
\endbibitem

\bibitem{2}
\begin{barticle}
\bauthor{\bsnm{{Maher}}, \binits{R.C.}}:
\batitle{Audio forensic examination}.
\bjtitle{IEEE Signal Processing Magazine.}
\bvolume{26},
\bfpage{84}--\blpage{94}
(\byear{2009})
\end{barticle}
\endbibitem

\bibitem{35}
\begin{barticle}
\bauthor{\bsnm{Steinebach}, \binits{M.}},
\bauthor{\bsnm{Dittmann}, \binits{J.}}:
\batitle{Watermarking-based digital audio data authentication}.
\bjtitle{EURASIP J. Adv. Signal Process.}
\bvolume{2003},
\bfpage{1001}--\blpage{1015}
(\byear{2003})
\end{barticle}
\endbibitem

\bibitem{3}
\begin{bchapter}
\bauthor{\bsnm{{Garcia-Romero}}, \binits{D.}},
\bauthor{\bsnm{{Espy-Wilson}}, \binits{C.Y.}}:
\bctitle{Automatic acquisition device identification from speech recordings}.
In: \bbtitle{2010 IEEE International Conference on Acoustics, Speech and Signal
  Processing},
pp. \bfpage{1806}--\blpage{1809}
(\byear{2010})
\end{bchapter}
\endbibitem

\bibitem{4}
\begin{bchapter}
\bauthor{\bsnm{{Hadoltikar}}, \binits{V.A.}},
\bauthor{\bsnm{{Ratnaparkhe}}, \binits{V.R.}},
\bauthor{\bsnm{{Kumar}}, \binits{R.}}:
\bctitle{Optimization of mfcc parameters for mobile phone recognition from
  audio recordings}.
In: \bbtitle{2019 3rd International Conference on Electronics, Communication
  and Aerospace Technology (ICECA)},
pp. \bfpage{777}--\blpage{780}
(\byear{2019})
\end{bchapter}
\endbibitem

\bibitem{Zeng2022a}
\begin{barticle}
\bauthor{\bsnm{Zeng}, \binits{C.}},
\bauthor{\bsnm{Yang}, \binits{Y.}},
\bauthor{\bsnm{Wang}, \binits{Z.}},
\bauthor{\bsnm{Kong}, \binits{S.}},
\bauthor{\bsnm{Feng}, \binits{S.}}:
\batitle{Audio tampering forensics based on representation learning of enf
  phase sequence}.
\bjtitle{International Journal of Digital Crime and Forensics}
\bvolume{14}(\bissue{1}),
\bfpage{1}--\blpage{19}
(\byear{2022}).
doi:\doiurl{10.4018/IJDCF.302894}
\end{barticle}
\endbibitem

\bibitem{Zeng2020}
\begin{barticle}
\bauthor{\bsnm{Zeng}, \binits{C.}},
\bauthor{\bsnm{Zhu}, \binits{D.}},
\bauthor{\bsnm{Wang}, \binits{Z.}},
\bauthor{\bsnm{Wang}, \binits{Z.}},
\bauthor{\bsnm{Zhao}, \binits{N.}},
\bauthor{\bsnm{He}, \binits{L.}}:
\batitle{An end-to-end deep source recording device identification system for
  web media forensics}.
\bjtitle{International Journal of Web Information Systems}
\bvolume{16}(\bissue{4}),
\bfpage{413}--\blpage{425}
(\byear{2020}).
doi:\doiurl{10.1108/IJWIS-06-2020-0038}
\end{barticle}
\endbibitem

\bibitem{Zeng2021b}
\begin{bchapter}
\bauthor{\bsnm{Zeng}, \binits{C.}},
\bauthor{\bsnm{Zhu}, \binits{D.}},
\bauthor{\bsnm{Wang}, \binits{Z.}},
\bauthor{\bsnm{Yang}, \binits{Y.}}:
\bctitle{Deep and shallow feature fusion and recognition of recording devices
  based on attention mechanism}.
In: \beditor{\bsnm{Barolli}, \binits{L.}},
\beditor{\bsnm{Li}, \binits{K.F.}},
\beditor{\bsnm{Miwa}, \binits{H.}} (eds.)
\bbtitle{Advances in Intelligent Networking and Collaborative Systems}
vol. \bseriesno{1263},
pp. \bfpage{372}--\blpage{381}.
\bpublisher{Springer},
\blocation{Cham}
(\byear{2021})
\end{bchapter}
\endbibitem

\bibitem{5}
\begin{bchapter}
\bauthor{\bsnm{{Zou}}, \binits{L.}},
\bauthor{\bsnm{{Yang}}, \binits{J.}},
\bauthor{\bsnm{{Huang}}, \binits{T.}}:
\bctitle{Automatic cell phone recognition from speech recordings}.
In: \bbtitle{2014 IEEE China Summit International Conference on Signal and
  Information Processing (ChinaSIP)},
pp. \bfpage{621}--\blpage{625}
(\byear{2014})
\end{bchapter}
\endbibitem

\bibitem{6}
\begin{barticle}
\bauthor{\bsnm{{Hanilci}}, \binits{C.}},
\bauthor{\bsnm{{Ertas}}, \binits{F.}},
\bauthor{\bsnm{{Ertas}}, \binits{T.}},
\bauthor{\bsnm{{Eskidere}}, \binits{A.}}:
\batitle{Recognition of brand and models of cell-phones from recorded speech
  signals}.
\bjtitle{IEEE Transactions on Information Forensics and Security.}
\bvolume{7},
\bfpage{625}--\blpage{634}
(\byear{2012})
\end{barticle}
\endbibitem

\bibitem{7}
\begin{botherref}
\oauthor{\bsnm{Hanilçi}, \binits{C.}},
\oauthor{\bsnm{Kinnunen}, \binits{T.}}:
Source cell-phone recognition from recorded speech using non-speech segments.
Digital Signal Processing.
\textbf{35}
(2014)
\end{botherref}
\endbibitem

\bibitem{8}
\begin{barticle}
\bauthor{\bsnm{Garcia-Romero}, \binits{D.}},
\bauthor{\bsnm{Espy-Wilson}, \binits{C.}}:
\batitle{Speech forensics: Automatic acquisition device identification}.
\bjtitle{The Journal of the Acoustical Society of America.}
\bvolume{127},
\bfpage{2044}
(\byear{2010})
\end{barticle}
\endbibitem

\bibitem{9}
\begin{bchapter}
\bauthor{\bsnm{{Kotropoulos}}, \binits{C.}},
\bauthor{\bsnm{{Samaras}}, \binits{S.}}:
\bctitle{Mobile phone identification using recorded speech signals}.
In: \bbtitle{2014 19th International Conference on Digital Signal Processing},
pp. \bfpage{586}--\blpage{591}
(\byear{2014})
\end{bchapter}
\endbibitem

\bibitem{Zeng2022}
\begin{barticle}
\bauthor{\bsnm{Zeng}, \binits{C.}},
\bauthor{\bsnm{Ye}, \binits{J.}},
\bauthor{\bsnm{Wang}, \binits{Z.}},
\bauthor{\bsnm{Zhao}, \binits{N.}},
\bauthor{\bsnm{Wu}, \binits{M.}}:
\batitle{Cascade neural network-based joint sampling and reconstruction for
  image compressed sensing}.
\bjtitle{Signal, Image and Video Processing}
\bvolume{16}(\bissue{1}),
\bfpage{47}--\blpage{54}
(\byear{2022}).
doi:\doiurl{10.1007/s11760-021-01955-w}
\end{barticle}
\endbibitem

\bibitem{10}
\begin{barticle}
\bauthor{\bsnm{{Baldini}}, \binits{G.}},
\bauthor{\bsnm{{Amerini}}, \binits{I.}}:
\batitle{Smartphones identification through the built-in microphones with
  convolutional neural network.}
\bjtitle{IEEE Access.}
\bvolume{7},
\bfpage{158685}--\blpage{158696}
(\byear{2019})
\end{barticle}
\endbibitem

\bibitem{Zeng2022b}
\begin{botherref}
\oauthor{\bsnm{Zeng}, \binits{C.}},
\oauthor{\bsnm{Yan}, \binits{K.}},
\oauthor{\bsnm{Wang}, \binits{Z.}},
\oauthor{\bsnm{Yu}, \binits{Y.}},
\oauthor{\bsnm{Xia}, \binits{S.}},
\oauthor{\bsnm{Zhao}, \binits{N.}}:
Abs-cam: A gradient optimization interpretable approach for explanation of
  convolutional neural networks.
Signal, Image and Video Processing,
1--8
(2022).
doi:\doiurl{10.1007/s11760-022-02313-0}
\end{botherref}
\endbibitem

\bibitem{11}
\begin{barticle}
\bauthor{\bsnm{{Li}}, \binits{Y.}},
\bauthor{\bsnm{{Zhang}}, \binits{X.}},
\bauthor{\bsnm{{Li}}, \binits{X.}},
\bauthor{\bsnm{{Zhang}}, \binits{Y.}},
\bauthor{\bsnm{{Yang}}, \binits{J.}},
\bauthor{\bsnm{{He}}, \binits{Q.}}:
\batitle{Mobile phone clustering from speech recordings using deep
  representation and spectral clustering}.
\bjtitle{IEEE Transactions on Information Forensics and Security.}
\bvolume{13},
\bfpage{965}--\blpage{977}
(\byear{2018})
\end{barticle}
\endbibitem

\bibitem{Wang2022ac}
\begin{botherref}
\oauthor{\bsnm{Wang}, \binits{Z.}},
\oauthor{\bsnm{Wang}, \binits{Z.}},
\oauthor{\bsnm{Zeng}, \binits{C.}},
\oauthor{\bsnm{Yu}, \binits{Y.}},
\oauthor{\bsnm{Wan}, \binits{X.}}:
High-quality image compressed sensing and reconstruction with multi-scale
  dilated convolutional neural network.
Circuits, Systems, and Signal Processing,
1--24
(2022).
doi:\doiurl{10.1007/s00034-022-02181-6}
\end{botherref}
\endbibitem

\bibitem{12}
\begin{barticle}
\bauthor{\bsnm{{Ashraf}}, \binits{M.}},
\bauthor{\bsnm{{Geng}}, \binits{G.}},
\bauthor{\bsnm{{Wang}}, \binits{X.}},
\bauthor{\bsnm{{Ahmad}}, \binits{F.}},
\bauthor{\bsnm{{Abid}}, \binits{F.}}:
\batitle{A globally regularized joint neural architecture for music
  classification}.
\bjtitle{IEEE Access.}
\bvolume{8},
\bfpage{220980}--\blpage{220989}
(\byear{2020})
\end{barticle}
\endbibitem

\bibitem{Lyu2022}
\begin{barticle}
\bauthor{\bsnm{Lyu}, \binits{L.}},
\bauthor{\bsnm{Wang}, \binits{Z.}},
\bauthor{\bsnm{Yun}, \binits{H.}},
\bauthor{\bsnm{Yang}, \binits{Z.}},
\bauthor{\bsnm{Li}, \binits{Y.}}:
\batitle{Deep knowledge tracing based on spatial and temporal representation
  learning for learning performance prediction}.
\bjtitle{Applied Sciences}
\bvolume{12}(\bissue{14}),
\bfpage{1}--\blpage{21}
(\byear{2022}).
doi:\doiurl{10.3390/app12147188}
\end{barticle}
\endbibitem

\bibitem{13}
\begin{botherref}
\oauthor{\bsnm{Zeng}, \binits{C.}},
\oauthor{\bsnm{Zhu}, \binits{D.}},
\oauthor{\bsnm{Wang}, \binits{Z.}},
\oauthor{\bsnm{Wang}, \binits{Z.}},
\oauthor{\bsnm{Zhao}, \binits{N.}},
\oauthor{\bsnm{He}, \binits{L.}}:
An end-to-end deep source recording device identification system for web media
  forensics.
International Journal of Web Information Systems.
(2020)
\end{botherref}
\endbibitem

\bibitem{Zeng2021c}
\begin{barticle}
\bauthor{\bsnm{Zeng}, \binits{C.}},
\bauthor{\bsnm{Wang}, \binits{Z.}},
\bauthor{\bsnm{Wang}, \binits{Z.}},
\bauthor{\bsnm{Yan}, \binits{K.}},
\bauthor{\bsnm{Yu}, \binits{Y.}}:
\batitle{Image compressed sensing and reconstruction of multi-scale residual
  network combined with channel attention mechanism}.
\bjtitle{Journal of Physics: Conference Series}
\bvolume{2010}(\bissue{1}),
\bfpage{012134}
(\byear{2021}).
doi:\doiurl{10.1088/1742-6596/2010/1/012134}
\end{barticle}
\endbibitem

\bibitem{Wang2021}
\begin{barticle}
\bauthor{\bsnm{Wang}, \binits{Z.}},
\bauthor{\bsnm{Zuo}, \binits{C.}},
\bauthor{\bsnm{Zeng}, \binits{C.}}:
\batitle{Sae based unified double jpeg compression detection system for web
  image forensics}.
\bjtitle{International Journal of Web Information Systems}
\bvolume{17}(\bissue{2}),
\bfpage{84}--\blpage{98}
(\byear{2021}).
doi:\doiurl{10.1108/IJWIS-11-2020-0073}
\end{barticle}
\endbibitem

\bibitem{Wang2021m}
\begin{bchapter}
\bauthor{\bsnm{Wang}, \binits{Z.}},
\bauthor{\bsnm{Zeng}, \binits{C.}},
\bauthor{\bsnm{Duan}, \binits{S.}},
\bauthor{\bsnm{Ouyang}, \binits{H.}},
\bauthor{\bsnm{Xu}, \binits{H.}}:
\bctitle{Robust speaker recognition based on stacked auto-encoders}.
In: \beditor{\bsnm{Barolli}, \binits{L.}},
\beditor{\bsnm{Li}, \binits{K.F.}},
\beditor{\bsnm{Enokido}, \binits{T.}},
\beditor{\bsnm{Takizawa}, \binits{M.}} (eds.)
\bbtitle{Advances in Networked-Based Information Systems}
vol. \bseriesno{1264},
pp. \bfpage{390}--\blpage{399}.
\bpublisher{Springer},
\blocation{Cham}
(\byear{2021})
\end{bchapter}
\endbibitem

\bibitem{Wang2020h}
\begin{bchapter}
\bauthor{\bsnm{Wang}, \binits{Z.}},
\bauthor{\bsnm{Duan}, \binits{S.}},
\bauthor{\bsnm{Zeng}, \binits{C.}},
\bauthor{\bsnm{Yu}, \binits{X.}},
\bauthor{\bsnm{Yang}, \binits{Y.}},
\bauthor{\bsnm{Wu}, \binits{H.}}:
\bctitle{Robust speaker identification of iot based on stacked sparse denoising
  auto-encoders}.
In: \bbtitle{2020 International Conferences on Internet of Things (iThings)},
pp. \bfpage{252}--\blpage{257}.
\bpublisher{IEEE},
\blocation{Rhodes, Greece}
(\byear{2020}).
doi:\doiurl{10.1109/iThings-GreenCom-CPSCom-SmartData-Cybermatics50389.2020.00056}
\end{bchapter}
\endbibitem

\bibitem{Zeng2018}
\begin{bchapter}
\bauthor{\bsnm{Zeng}, \binits{C.-Y.}},
\bauthor{\bsnm{Ma}, \binits{C.-F.}},
\bauthor{\bsnm{Wang}, \binits{Z.-F.}},
\bauthor{\bsnm{Ye}, \binits{J.-X.}}:
\bctitle{Stacked autoencoder networks based speaker recognition}.
In: \bbtitle{2018 International Conference on Machine Learning and Cybernetics
  (ICMLC)},
pp. \bfpage{294}--\blpage{299}.
\bpublisher{IEEE},
\blocation{Chengdu}
(\byear{2018}).
doi:\doiurl{10.1109/ICMLC.2018.8526953}
\end{bchapter}
\endbibitem

\bibitem{Wang2015b}
\begin{bchapter}
\bauthor{\bsnm{Wang}, \binits{Z.}},
\bauthor{\bsnm{Liu}, \binits{Q.}},
\bauthor{\bsnm{Chen}, \binits{J.}},
\bauthor{\bsnm{Yao}, \binits{H.}}:
\bctitle{Recording source identification using device universal background
  model}.
In: \bbtitle{2015 International Conference of Educational Innovation Through
  Technology (EITT)},
pp. \bfpage{19}--\blpage{23}.
\bpublisher{IEEE},
\blocation{Wuhan, China}
(\byear{2015}).
doi:\doiurl{10.1109/EITT.2015.11}
\end{bchapter}
\endbibitem

\bibitem{Zeng2020a}
\begin{bchapter}
\bauthor{\bsnm{Zeng}, \binits{C.}},
\bauthor{\bsnm{Wang}, \binits{Z.}},
\bauthor{\bsnm{Wang}, \binits{Z.}}:
\bctitle{Image reconstruction of iot based on parallel cnn}.
In: \bbtitle{2020 International Conferences on Internet of Things (iThings)},
pp. \bfpage{258}--\blpage{263}.
\bpublisher{IEEE},
\blocation{Rhodes, Greece}
(\byear{2020}).
doi:\doiurl{10.1109/iThings-GreenCom-CPSCom-SmartData-Cybermatics50389.2020.00057}
\end{bchapter}
\endbibitem

\bibitem{Wang2018a}
\begin{bchapter}
\bauthor{\bsnm{Wang}, \binits{Z.-F.}},
\bauthor{\bsnm{Wang}, \binits{J.}},
\bauthor{\bsnm{Zeng}, \binits{C.-Y.}},
\bauthor{\bsnm{Min}, \binits{Q.-S.}},
\bauthor{\bsnm{Tian}, \binits{Y.}},
\bauthor{\bsnm{Zuo}, \binits{M.-Z.}}:
\bctitle{Digital audio tampering detection based on enf consistency}.
In: \bbtitle{2018 International Conference on Wavelet Analysis and Pattern
  Recognition (ICWAPR)},
pp. \bfpage{209}--\blpage{214}.
\bpublisher{IEEE},
\blocation{Chengdu}
(\byear{2018}).
doi:\doiurl{10.1109/ICWAPR.2018.8521378}
\end{bchapter}
\endbibitem

\bibitem{Wang2017}
\begin{bchapter}
\bauthor{\bsnm{Wang}, \binits{Z.-F.}},
\bauthor{\bsnm{Zhu}, \binits{L.}},
\bauthor{\bsnm{Min}, \binits{Q.-S.}},
\bauthor{\bsnm{Zeng}, \binits{C.-Y.}}:
\bctitle{Double compression detection based on feature fusion}.
In: \bbtitle{2017 International Conference on Machine Learning and Cybernetics
  (ICMLC)},
pp. \bfpage{379}--\blpage{384}.
\bpublisher{IEEE},
\blocation{Ningbo}
(\byear{2017}).
doi:\doiurl{10.1109/ICMLC.2017.8108951}
\end{bchapter}
\endbibitem

\bibitem{Wang2015a}
\begin{bchapter}
\bauthor{\bsnm{Wang}, \binits{Z.}},
\bauthor{\bsnm{Liu}, \binits{Q.}},
\bauthor{\bsnm{Yao}, \binits{H.}},
\bauthor{\bsnm{Chen}, \binits{J.}}:
\bctitle{Virtual chime-bells experimental system based on multi-modal fusion}.
In: \bbtitle{2015 International Conference of Educational Innovation Through
  Technology (EITT)},
pp. \bfpage{64}--\blpage{67}.
\bpublisher{IEEE},
\blocation{Wuhan, China}
(\byear{2015}).
doi:\doiurl{10.1109/EITT.2015.20}
\end{bchapter}
\endbibitem

\bibitem{14}
\begin{bchapter}
\bauthor{\bsnm{Kraetzer}, \binits{C.}},
\bauthor{\bsnm{Oermann}, \binits{A.}},
\bauthor{\bsnm{Dittmann}, \binits{J.}},
\bauthor{\bsnm{Lang}, \binits{A.}}:
\bctitle{Digital audio forensics: A first practical evaluation on microphone
  and environment classification}.
In: \bbtitle{Workshop on Multimedia \& Security},
pp. \bfpage{63}--\blpage{74}
(\byear{2007})
\end{bchapter}
\endbibitem

\bibitem{15}
\begin{barticle}
\bauthor{\bsnm{Qin}, \binits{T.}},
\bauthor{\bsnm{Wang}, \binits{R.}},
\bauthor{\bsnm{Yan}, \binits{D.}},
\bauthor{\bsnm{Lin}, \binits{L.}}:
\batitle{Source cell-phone identification in the presence of additive noise
  from cqt domain}.
\bjtitle{Information(Switzerland).}
\bvolume{9},
\bfpage{205}
(\byear{2018})
\end{barticle}
\endbibitem

\bibitem{16}
\begin{botherref}
\oauthor{\bsnm{Eskidere}, \binits{A.}}:
Source digital voice recorder identification by wavelet analysis.
International Journal on Artificial Intelligence Tools.
\textbf{25}
(2016)
\end{botherref}
\endbibitem

\bibitem{17}
\begin{bchapter}
\bauthor{\bsnm{Hanilçi}, \binits{C.}},
\bauthor{\bsnm{Ertas}, \binits{F.}}:
\bctitle{Optimizing acoustic features for source cell-phone recognition using
  speech signals}.
In: \bbtitle{Acm Workshop on Information Hiding \& Multimedia Security},
pp. \bfpage{141}--\blpage{148}
(\byear{2013})
\end{bchapter}
\endbibitem

\bibitem{18}
\begin{bchapter}
\bauthor{\bsnm{{Aggarwal}}, \binits{R.}},
\bauthor{\bsnm{{Singh}}, \binits{S.}},
\bauthor{\bsnm{{Roul}}, \binits{A.K.}},
\bauthor{\bsnm{{Khanna}}, \binits{N.}}:
\bctitle{Cellphone identification using noise estimates from recorded audio}.
In: \bbtitle{2014 International Conference on Communication and Signal
  Processing},
pp. \bfpage{1218}--\blpage{1222}
(\byear{2014})
\end{bchapter}
\endbibitem

\bibitem{19}
\begin{bchapter}
\bauthor{\bsnm{{Zou}}, \binits{L.}},
\bauthor{\bsnm{{He}}, \binits{Q.}},
\bauthor{\bsnm{{Yang}}, \binits{J.}},
\bauthor{\bsnm{{Li}}, \binits{Y.}}:
\bctitle{Source cell phone matching from speech recordings by sparse
  representation and kiss metric}.
In: \bbtitle{2016 IEEE International Conference on Acoustics, Speech and Signal
  Processing (ICASSP)},
pp. \bfpage{2079}--\blpage{2083}
(\byear{2016})
\end{bchapter}
\endbibitem

\bibitem{20}
\begin{barticle}
\bauthor{\bsnm{Jin}, \binits{C.}},
\bauthor{\bsnm{Wang}, \binits{R.}},
\bauthor{\bsnm{Yan}, \binits{D.}}:
\batitle{Source smartphone identification by exploiting encoding
  characteristics of recorded speech}.
\bjtitle{Digital Investigation.}
\bvolume{29},
\bfpage{129}--\blpage{146}
(\byear{2019})
\end{barticle}
\endbibitem

\bibitem{21}
\begin{barticle}
\bauthor{\bsnm{{Campbell}}, \binits{W.M.}},
\bauthor{\bsnm{{Sturim}}, \binits{D.E.}},
\bauthor{\bsnm{{Reynolds}}, \binits{D.A.}}:
\batitle{Support vector machines using gmm supervectors for speaker
  verification}.
\bjtitle{IEEE Signal Processing Letters.}
\bvolume{13},
\bfpage{308}--\blpage{311}
(\byear{2006})
\end{barticle}
\endbibitem

\bibitem{22}
\begin{barticle}
\bauthor{\bsnm{{Jiang}}, \binits{Y.}},
\bauthor{\bsnm{{Leung}}, \binits{F.H.F.}}:
\batitle{Source microphone recognition aided by a kernel-based projection
  method}.
\bjtitle{IEEE Transactions on Information Forensics and Security.}
\bvolume{14},
\bfpage{2875}--\blpage{2886}
(\byear{2019})
\end{barticle}
\endbibitem

\bibitem{23}
\begin{bchapter}
\bauthor{\bsnm{{Li}}, \binits{Y.}},
\bauthor{\bsnm{{Zhang}}, \binits{X.}},
\bauthor{\bsnm{{Li}}, \binits{X.}},
\bauthor{\bsnm{{Feng}}, \binits{X.}},
\bauthor{\bsnm{{Yang}}, \binits{J.}},
\bauthor{\bsnm{{Chen}}, \binits{A.}},
\bauthor{\bsnm{{He}}, \binits{Q.}}:
\bctitle{Mobile phone clustering from acquired speech recordings using deep
  gaussian supervector and spectral clustering}.
In: \bbtitle{2017 IEEE International Conference on Acoustics, Speech and Signal
  Processing (ICASSP)},
pp. \bfpage{2137}--\blpage{2141}
(\byear{2017})
\end{bchapter}
\endbibitem

\bibitem{24}
\begin{barticle}
\bauthor{\bsnm{{Lin}}, \binits{X.}},
\bauthor{\bsnm{{Zhu}}, \binits{J.}},
\bauthor{\bsnm{{Chen}}, \binits{D.}}:
\batitle{Subband aware cnn for cell-phone recognition}.
\bjtitle{IEEE Signal Processing Letters.}
\bvolume{27},
\bfpage{605}--\blpage{609}
(\byear{2020})
\end{barticle}
\endbibitem

\bibitem{25}
\begin{bchapter}
\bauthor{\bsnm{{Campbell}}, \binits{W.M.}}:
\bctitle{Generalized linear discriminant sequence kernels for speaker
  recognition}.
In: \bbtitle{2002 IEEE International Conference on Acoustics, Speech, and
  Signal Processing},
vol. \bseriesno{1},
pp. \bfpage{161}--\blpage{164}
(\byear{2002})
\end{bchapter}
\endbibitem

\bibitem{26}
\begin{barticle}
\bauthor{\bsnm{{Baldini}}, \binits{G.}},
\bauthor{\bsnm{{Amerini}}, \binits{I.}},
\bauthor{\bsnm{{Gentile}}, \binits{C.}}:
\batitle{Microphone identification using convolutional neural networks}.
\bjtitle{IEEE Sensors Letters.}
\bvolume{3},
\bfpage{1}--\blpage{4}
(\byear{2019})
\end{barticle}
\endbibitem

\bibitem{27}
\begin{barticle}
\bauthor{\bsnm{{Baldini}}, \binits{G.}},
\bauthor{\bsnm{{Steri}}, \binits{G.}}:
\batitle{A survey of techniques for the identification of mobile phones using
  the physical fingerprints of the built-in components}.
\bjtitle{IEEE Communications Surveys Tutorials.}
\bvolume{19},
\bfpage{1761}--\blpage{1789}
(\byear{2017})
\end{barticle}
\endbibitem

\bibitem{28}
\begin{barticle}
\bauthor{\bsnm{{Luo}}, \binits{D.}},
\bauthor{\bsnm{{Korus}}, \binits{P.}},
\bauthor{\bsnm{{Huang}}, \binits{J.}}:
\batitle{Band energy difference for source attribution in audio forensics}.
\bjtitle{IEEE Transactions on Information Forensics and Security.}
\bvolume{13},
\bfpage{2179}--\blpage{2189}
(\byear{2018})
\end{barticle}
\endbibitem

\bibitem{29}
\begin{barticle}
\bauthor{\bsnm{Zou}, \binits{L.}},
\bauthor{\bsnm{He}, \binits{Q.}},
\bauthor{\bsnm{Wu}, \binits{J.}}:
\batitle{Source cell phone verification from speech recordings using sparse
  representation}.
\bjtitle{Digital Signal Processing.}
\bvolume{62},
\bfpage{125}--\blpage{136}
(\byear{2016})
\end{barticle}
\endbibitem

\bibitem{30}
\begin{bchapter}
\bauthor{\bsnm{{Zou}}, \binits{L.}},
\bauthor{\bsnm{{He}}, \binits{Q.}},
\bauthor{\bsnm{{Feng}}, \binits{X.}}:
\bctitle{Cell phone verification from speech recordings using sparse
  representation}.
In: \bbtitle{2015 IEEE International Conference on Acoustics, Speech and Signal
  Processing (ICASSP)},
pp. \bfpage{1787}--\blpage{1791}
(\byear{2015})
\end{bchapter}
\endbibitem

\bibitem{31}
\begin{barticle}
\bauthor{\bsnm{{Aharon}}, \binits{M.}},
\bauthor{\bsnm{{Elad}}, \binits{M.}},
\bauthor{\bsnm{{Bruckstein}}, \binits{A.}}:
\batitle{K-svd: An algorithm for designing overcomplete dictionaries for sparse
  representation}.
\bjtitle{IEEE Transactions on Signal Processing.}
\bvolume{54},
\bfpage{4311}--\blpage{4322}
(\byear{2006})
\end{barticle}
\endbibitem

\bibitem{32}
\begin{bchapter}
\bauthor{\bsnm{{Zhang}}, \binits{Q.}},
\bauthor{\bsnm{{Li}}, \binits{B.}}:
\bctitle{Discriminative k-svd for dictionary learning in face recognition}.
In: \bbtitle{2010 IEEE Computer Society Conference on Computer Vision and
  Pattern Recognition},
pp. \bfpage{2691}--\blpage{2698}
(\byear{2010})
\end{bchapter}
\endbibitem

\bibitem{36}
\begin{botherref}
\oauthor{\bsnm{Pan}, \binits{X.}},
\oauthor{\bsnm{Shi}, \binits{J.}},
\oauthor{\bsnm{Luo}, \binits{P.}},
\oauthor{\bsnm{Wang}, \binits{X.}},
\oauthor{\bsnm{Tang}, \binits{X.}}:
Spatial as deep: Spatial cnn for traffic scene understanding
(2017)
\end{botherref}
\endbibitem

\bibitem{33}
\begin{bchapter}
\bauthor{\bsnm{{He}}, \binits{K.}},
\bauthor{\bsnm{{Zhang}}, \binits{X.}},
\bauthor{\bsnm{{Ren}}, \binits{S.}},
\bauthor{\bsnm{{Sun}}, \binits{J.}}:
\bctitle{Deep residual learning for image recognition}.
In: \bbtitle{2016 IEEE Conference on Computer Vision and Pattern Recognition
  (CVPR)},
pp. \bfpage{770}--\blpage{778}
(\byear{2016})
\end{bchapter}
\endbibitem

\bibitem{34}
\begin{barticle}
\bauthor{\bsnm{{Xie}}, \binits{Y.}},
\bauthor{\bsnm{{Liang}}, \binits{R.}},
\bauthor{\bsnm{{Liang}}, \binits{Z.}},
\bauthor{\bsnm{{Huang}}, \binits{C.}},
\bauthor{\bsnm{{Zou}}, \binits{C.}},
\bauthor{\bsnm{{Schuller}}, \binits{B.}}:
\batitle{Speech emotion classification using attention-based lstm}.
\bjtitle{IEEE/ACM Transactions on Audio, Speech, and Language Processing.}
\bvolume{27},
\bfpage{1675}--\blpage{1685}
(\byear{2019})
\end{barticle}
\endbibitem

\bibitem{37}
\begin{barticle}
\bauthor{\bsnm{Campbell}, \binits{W.M.}},
\bauthor{\bsnm{Sturim}, \binits{D.E.}},
\bauthor{\bsnm{Reynolds}, \binits{D.A.}}:
\batitle{Support vector machines using gmm supervectors for speaker
  verification}.
\bjtitle{IEEE Signal Processing Letters}
\bvolume{13}(\bissue{5}),
\bfpage{308}--\blpage{311}
(\byear{2006}).
doi:\doiurl{10.1109/LSP.2006.870086}
\end{barticle}
\endbibitem

\end{thebibliography}

\newcommand{\BMCxmlcomment}[1]{}

\BMCxmlcomment{

<refgrp>

<bibl id="B1">
  <title><p>Spatial and Temporal Learning Representation for End-to-End
  Recording Device Identification</p></title>
  <aug>
    <au><snm>Zeng</snm><fnm>C</fnm></au>
    <au><snm>Zhu</snm><fnm>D</fnm></au>
    <au><snm>Wang</snm><fnm>Z</fnm></au>
    <au><snm>Wu</snm><fnm>M</fnm></au>
    <au><snm>Xiong</snm><fnm>W</fnm></au>
    <au><snm>Zhao</snm><fnm>N</fnm></au>
  </aug>
  <source>EURASIP Journal on Advances in Signal Processing</source>
  <pubdate>2021</pubdate>
  <volume>2021</volume>
  <issue>1</issue>
  <fpage>41</fpage>
</bibl>

<bibl id="B2">
  <title><p>Shallow and Deep Feature Fusion for Digital Audio Tampering
  Detection</p></title>
  <aug>
    <au><snm>Wang</snm><fnm>Z</fnm></au>
    <au><snm>Yang</snm><fnm>Y</fnm></au>
    <au><snm>Zeng</snm><fnm>C</fnm></au>
    <au><snm>Kong</snm><fnm>S</fnm></au>
    <au><snm>Feng</snm><fnm>S</fnm></au>
    <au><snm>Zhao</snm><fnm>N</fnm></au>
  </aug>
  <source>EURASIP Journal on Advances in Signal Processing</source>
  <pubdate>2022</pubdate>
  <volume>2022</volume>
  <issue>69</issue>
  <fpage>1</fpage>
  <lpage>-20</lpage>
</bibl>

<bibl id="B3">
  <title><p>Acoustic Scene Identification for Audio Authentication</p></title>
  <aug>
    <au><snm>Narkhede</snm><fnm>M</fnm></au>
    <au><snm>Patole</snm><fnm>R</fnm></au>
  </aug>
  <source>Soft Computing and Signal Processing</source>
  <editor>Wang, Jiacun and Reddy, G. Ram Mohana and Prasad, V. Kamakshi and
  Reddy, V. Sivakumar</editor>
  <pubdate>2019</pubdate>
  <fpage>593</fpage>
  <lpage>602</lpage>
</bibl>

<bibl id="B4">
  <title><p>Audio forensic examination</p></title>
  <aug>
    <au><snm>{Maher}</snm><fnm>R. C.</fnm></au>
  </aug>
  <source>IEEE Signal Processing Magazine.</source>
  <pubdate>2009</pubdate>
  <volume>26</volume>
  <fpage>84</fpage>
  <lpage>94</lpage>
</bibl>

<bibl id="B5">
  <title><p>Watermarking-Based Digital Audio Data Authentication</p></title>
  <aug>
    <au><snm>Steinebach</snm><fnm>M</fnm></au>
    <au><snm>Dittmann</snm><fnm>J</fnm></au>
  </aug>
  <source>EURASIP J. Adv. Signal Process.</source>
  <pubdate>2003</pubdate>
  <volume>2003</volume>
  <fpage>1001–1015</fpage>
</bibl>

<bibl id="B6">
  <title><p>Automatic acquisition device identification from speech
  recordings</p></title>
  <aug>
    <au><snm>{Garcia-Romero}</snm><fnm>D.</fnm></au>
    <au><snm>{Espy-Wilson}</snm><fnm>C. Y.</fnm></au>
  </aug>
  <source>2010 IEEE International Conference on Acoustics, Speech and Signal
  Processing</source>
  <pubdate>2010</pubdate>
  <fpage>1806</fpage>
  <lpage>1809</lpage>
</bibl>

<bibl id="B7">
  <title><p>Optimization of MFCC parameters for mobile phone recognition from
  audio recordings</p></title>
  <aug>
    <au><snm>{Hadoltikar}</snm><fnm>V. A.</fnm></au>
    <au><snm>{Ratnaparkhe}</snm><fnm>V. R.</fnm></au>
    <au><snm>{Kumar}</snm><fnm>R.</fnm></au>
  </aug>
  <source>2019 3rd International conference on Electronics, Communication and
  Aerospace Technology (ICECA)</source>
  <pubdate>2019</pubdate>
  <fpage>777</fpage>
  <lpage>780</lpage>
</bibl>

<bibl id="B8">
  <title><p>Audio Tampering Forensics Based on Representation Learning of ENF
  Phase Sequence</p></title>
  <aug>
    <au><snm>Zeng</snm><fnm>C</fnm></au>
    <au><snm>Yang</snm><fnm>Y</fnm></au>
    <au><snm>Wang</snm><fnm>Z</fnm></au>
    <au><snm>Kong</snm><fnm>S</fnm></au>
    <au><snm>Feng</snm><fnm>S</fnm></au>
  </aug>
  <source>International Journal of Digital Crime and Forensics</source>
  <pubdate>2022</pubdate>
  <volume>14</volume>
  <issue>1</issue>
  <fpage>1</fpage>
  <lpage>-19</lpage>
</bibl>

<bibl id="B9">
  <title><p>An End-to-End Deep Source Recording Device Identification System
  for Web Media Forensics</p></title>
  <aug>
    <au><snm>Zeng</snm><fnm>C</fnm></au>
    <au><snm>Zhu</snm><fnm>D</fnm></au>
    <au><snm>Wang</snm><fnm>Z</fnm></au>
    <au><snm>Wang</snm><fnm>Z</fnm></au>
    <au><snm>Zhao</snm><fnm>N</fnm></au>
    <au><snm>He</snm><fnm>L</fnm></au>
  </aug>
  <source>International Journal of Web Information Systems</source>
  <pubdate>2020</pubdate>
  <volume>16</volume>
  <issue>4</issue>
  <fpage>413</fpage>
  <lpage>-425</lpage>
</bibl>

<bibl id="B10">
  <title><p>Deep and Shallow Feature Fusion and Recognition of Recording
  Devices Based on Attention Mechanism</p></title>
  <aug>
    <au><snm>Zeng</snm><fnm>C</fnm></au>
    <au><snm>Zhu</snm><fnm>D</fnm></au>
    <au><snm>Wang</snm><fnm>Z</fnm></au>
    <au><snm>Yang</snm><fnm>Y</fnm></au>
  </aug>
  <source>Advances in Intelligent Networking and Collaborative Systems</source>
  <publisher>Cham: Springer International Publishing</publisher>
  <editor>Barolli, Leonard and Li, Kin Fun and Miwa, Hiroyoshi</editor>
  <pubdate>2021</pubdate>
  <volume>1263</volume>
  <fpage>372</fpage>
  <lpage>-381</lpage>
</bibl>

<bibl id="B11">
  <title><p>Automatic cell phone recognition from speech recordings</p></title>
  <aug>
    <au><snm>{Zou}</snm><fnm>L.</fnm></au>
    <au><snm>{Yang}</snm><fnm>J.</fnm></au>
    <au><snm>{Huang}</snm><fnm>T.</fnm></au>
  </aug>
  <source>2014 IEEE China Summit International Conference on Signal and
  Information Processing (ChinaSIP)</source>
  <pubdate>2014</pubdate>
  <fpage>621</fpage>
  <lpage>625</lpage>
</bibl>

<bibl id="B12">
  <title><p>Recognition of Brand and Models of Cell-Phones From Recorded Speech
  Signals</p></title>
  <aug>
    <au><snm>{Hanilci}</snm><fnm>C.</fnm></au>
    <au><snm>{Ertas}</snm><fnm>F.</fnm></au>
    <au><snm>{Ertas}</snm><fnm>T.</fnm></au>
    <au><snm>{Eskidere}</snm><fnm>A.</fnm></au>
  </aug>
  <source>IEEE Transactions on Information Forensics and Security.</source>
  <pubdate>2012</pubdate>
  <volume>7</volume>
  <fpage>625</fpage>
  <lpage>634</lpage>
</bibl>

<bibl id="B13">
  <title><p>Source cell-phone recognition from recorded speech using non-speech
  segments</p></title>
  <aug>
    <au><snm>Hanilçi</snm><fnm>C</fnm></au>
    <au><snm>Kinnunen</snm><fnm>T</fnm></au>
  </aug>
  <source>Digital Signal Processing.</source>
  <pubdate>2014</pubdate>
  <volume>35</volume>
</bibl>

<bibl id="B14">
  <title><p>Speech forensics: Automatic acquisition device
  identification</p></title>
  <aug>
    <au><snm>Garcia Romero</snm><fnm>D</fnm></au>
    <au><snm>Espy Wilson</snm><fnm>C</fnm></au>
  </aug>
  <source>The Journal of the Acoustical Society of America.</source>
  <pubdate>2010</pubdate>
  <volume>127</volume>
  <fpage>2044</fpage>
</bibl>

<bibl id="B15">
  <title><p>Mobile phone identification using recorded speech
  signals</p></title>
  <aug>
    <au><snm>{Kotropoulos}</snm><fnm>C.</fnm></au>
    <au><snm>{Samaras}</snm><fnm>S.</fnm></au>
  </aug>
  <source>2014 19th International Conference on Digital Signal
  Processing</source>
  <pubdate>2014</pubdate>
  <fpage>586</fpage>
  <lpage>591</lpage>
</bibl>

<bibl id="B16">
  <title><p>Cascade Neural Network-Based Joint Sampling and Reconstruction for
  Image Compressed Sensing</p></title>
  <aug>
    <au><snm>Zeng</snm><fnm>C</fnm></au>
    <au><snm>Ye</snm><fnm>J</fnm></au>
    <au><snm>Wang</snm><fnm>Z</fnm></au>
    <au><snm>Zhao</snm><fnm>N</fnm></au>
    <au><snm>Wu</snm><fnm>M</fnm></au>
  </aug>
  <source>Signal, Image and Video Processing</source>
  <pubdate>2022</pubdate>
  <volume>16</volume>
  <issue>1</issue>
  <fpage>47</fpage>
  <lpage>-54</lpage>
</bibl>

<bibl id="B17">
  <title><p>Smartphones Identification Through the Built-In Microphones With
  Convolutional Neural Network.</p></title>
  <aug>
    <au><snm>{Baldini}</snm><fnm>G.</fnm></au>
    <au><snm>{Amerini}</snm><fnm>I.</fnm></au>
  </aug>
  <source>IEEE Access.</source>
  <pubdate>2019</pubdate>
  <volume>7</volume>
  <fpage>158685</fpage>
  <lpage>158696</lpage>
</bibl>

<bibl id="B18">
  <title><p>Abs-CAM: A Gradient Optimization Interpretable Approach for
  Explanation of Convolutional Neural Networks</p></title>
  <aug>
    <au><snm>Zeng</snm><fnm>C</fnm></au>
    <au><snm>Yan</snm><fnm>K</fnm></au>
    <au><snm>Wang</snm><fnm>Z</fnm></au>
    <au><snm>Yu</snm><fnm>Y</fnm></au>
    <au><snm>Xia</snm><fnm>S</fnm></au>
    <au><snm>Zhao</snm><fnm>N</fnm></au>
  </aug>
  <source>Signal, Image and Video Processing</source>
  <pubdate>2022</pubdate>
  <fpage>1</fpage>
  <lpage>-8</lpage>
</bibl>

<bibl id="B19">
  <title><p>Mobile Phone Clustering From Speech Recordings Using Deep
  Representation and Spectral Clustering</p></title>
  <aug>
    <au><snm>{Li}</snm><fnm>Y.</fnm></au>
    <au><snm>{Zhang}</snm><fnm>X.</fnm></au>
    <au><snm>{Li}</snm><fnm>X.</fnm></au>
    <au><snm>{Zhang}</snm><fnm>Y.</fnm></au>
    <au><snm>{Yang}</snm><fnm>J.</fnm></au>
    <au><snm>{He}</snm><fnm>Q.</fnm></au>
  </aug>
  <source>IEEE Transactions on Information Forensics and Security.</source>
  <pubdate>2018</pubdate>
  <volume>13</volume>
  <fpage>965</fpage>
  <lpage>977</lpage>
</bibl>

<bibl id="B20">
  <title><p>High-Quality Image Compressed Sensing and Reconstruction with
  Multi-Scale Dilated Convolutional Neural Network</p></title>
  <aug>
    <au><snm>Wang</snm><fnm>Z</fnm></au>
    <au><snm>Wang</snm><fnm>Z</fnm></au>
    <au><snm>Zeng</snm><fnm>C</fnm></au>
    <au><snm>Yu</snm><fnm>Y</fnm></au>
    <au><snm>Wan</snm><fnm>X</fnm></au>
  </aug>
  <source>Circuits, Systems, and Signal Processing</source>
  <pubdate>2022</pubdate>
  <fpage>1</fpage>
  <lpage>-24</lpage>
</bibl>

<bibl id="B21">
  <title><p>A Globally Regularized Joint Neural Architecture for Music
  Classification</p></title>
  <aug>
    <au><snm>{Ashraf}</snm><fnm>M.</fnm></au>
    <au><snm>{Geng}</snm><fnm>G.</fnm></au>
    <au><snm>{Wang}</snm><fnm>X.</fnm></au>
    <au><snm>{Ahmad}</snm><fnm>F.</fnm></au>
    <au><snm>{Abid}</snm><fnm>F.</fnm></au>
  </aug>
  <source>IEEE Access.</source>
  <pubdate>2020</pubdate>
  <volume>8</volume>
  <fpage>220980</fpage>
  <lpage>220989</lpage>
</bibl>

<bibl id="B22">
  <title><p>Deep Knowledge Tracing Based on Spatial and Temporal Representation
  Learning for Learning Performance Prediction</p></title>
  <aug>
    <au><snm>Lyu</snm><fnm>L</fnm></au>
    <au><snm>Wang</snm><fnm>Z</fnm></au>
    <au><snm>Yun</snm><fnm>H</fnm></au>
    <au><snm>Yang</snm><fnm>Z</fnm></au>
    <au><snm>Li</snm><fnm>Y</fnm></au>
  </aug>
  <source>Applied Sciences</source>
  <publisher>Multidisciplinary Digital Publishing Institute</publisher>
  <pubdate>2022</pubdate>
  <volume>12</volume>
  <issue>14</issue>
  <fpage>1</fpage>
  <lpage>-21</lpage>
</bibl>

<bibl id="B23">
  <title><p>An end-to-end deep source recording device identification system
  for Web media forensics</p></title>
  <aug>
    <au><snm>Zeng</snm><fnm>C</fnm></au>
    <au><snm>Zhu</snm><fnm>D</fnm></au>
    <au><snm>Wang</snm><fnm>Z</fnm></au>
    <au><snm>Wang</snm><fnm>Z</fnm></au>
    <au><snm>Zhao</snm><fnm>N</fnm></au>
    <au><snm>He</snm><fnm>L</fnm></au>
  </aug>
  <source>International Journal of Web Information Systems.</source>
  <pubdate>2020</pubdate>
</bibl>

<bibl id="B24">
  <title><p>Image Compressed Sensing and Reconstruction of Multi-Scale Residual
  Network Combined with Channel Attention Mechanism</p></title>
  <aug>
    <au><snm>Zeng</snm><fnm>C</fnm></au>
    <au><snm>Wang</snm><fnm>Z</fnm></au>
    <au><snm>Wang</snm><fnm>Z</fnm></au>
    <au><snm>Yan</snm><fnm>K</fnm></au>
    <au><snm>Yu</snm><fnm>Y</fnm></au>
  </aug>
  <source>Journal of Physics: Conference Series</source>
  <pubdate>2021</pubdate>
  <volume>2010</volume>
  <issue>1</issue>
  <fpage>012134</fpage>
</bibl>

<bibl id="B25">
  <title><p>SAE Based Unified Double JPEG Compression Detection System for Web
  Image Forensics</p></title>
  <aug>
    <au><snm>Wang</snm><fnm>Z</fnm></au>
    <au><snm>Zuo</snm><fnm>C</fnm></au>
    <au><snm>Zeng</snm><fnm>C</fnm></au>
  </aug>
  <source>International Journal of Web Information Systems</source>
  <pubdate>2021</pubdate>
  <volume>17</volume>
  <issue>2</issue>
  <fpage>84</fpage>
  <lpage>-98</lpage>
</bibl>

<bibl id="B26">
  <title><p>Robust Speaker Recognition Based on Stacked
  Auto-Encoders</p></title>
  <aug>
    <au><snm>Wang</snm><fnm>Z</fnm></au>
    <au><snm>Zeng</snm><fnm>C</fnm></au>
    <au><snm>Duan</snm><fnm>S</fnm></au>
    <au><snm>Ouyang</snm><fnm>H</fnm></au>
    <au><snm>Xu</snm><fnm>H</fnm></au>
  </aug>
  <source>Advances in Networked-Based Information Systems</source>
  <publisher>Cham: Springer International Publishing</publisher>
  <editor>Barolli, Leonard and Li, Kin Fun and Enokido, Tomoya and Takizawa,
  Makoto</editor>
  <pubdate>2021</pubdate>
  <volume>1264</volume>
  <fpage>390</fpage>
  <lpage>-399</lpage>
</bibl>

<bibl id="B27">
  <title><p>Robust Speaker Identification of IoT Based on Stacked Sparse
  Denoising Auto-Encoders</p></title>
  <aug>
    <au><snm>Wang</snm><fnm>Z</fnm></au>
    <au><snm>Duan</snm><fnm>S</fnm></au>
    <au><snm>Zeng</snm><fnm>C</fnm></au>
    <au><snm>Yu</snm><fnm>X</fnm></au>
    <au><snm>Yang</snm><fnm>Y</fnm></au>
    <au><snm>Wu</snm><fnm>H</fnm></au>
  </aug>
  <source>2020 International Conferences on Internet of Things
  (iThings)</source>
  <publisher>Rhodes, Greece: IEEE</publisher>
  <pubdate>2020</pubdate>
  <fpage>252</fpage>
  <lpage>-257</lpage>
</bibl>

<bibl id="B28">
  <title><p>Stacked Autoencoder Networks Based Speaker Recognition</p></title>
  <aug>
    <au><snm>Zeng</snm><fnm>CY</fnm></au>
    <au><snm>Ma</snm><fnm>CF</fnm></au>
    <au><snm>Wang</snm><fnm>ZF</fnm></au>
    <au><snm>Ye</snm><fnm>JX</fnm></au>
  </aug>
  <source>2018 International Conference on Machine Learning and Cybernetics
  (ICMLC)</source>
  <publisher>Chengdu: IEEE</publisher>
  <pubdate>2018</pubdate>
  <fpage>294</fpage>
  <lpage>-299</lpage>
</bibl>

<bibl id="B29">
  <title><p>Recording Source Identification Using Device Universal Background
  Model</p></title>
  <aug>
    <au><snm>Wang</snm><fnm>Z</fnm></au>
    <au><snm>Liu</snm><fnm>Q</fnm></au>
    <au><snm>Chen</snm><fnm>J</fnm></au>
    <au><snm>Yao</snm><fnm>H</fnm></au>
  </aug>
  <source>2015 International Conference of Educational Innovation through
  Technology (EITT)</source>
  <publisher>Wuhan, China: IEEE</publisher>
  <pubdate>2015</pubdate>
  <fpage>19</fpage>
  <lpage>-23</lpage>
</bibl>

<bibl id="B30">
  <title><p>Image Reconstruction of IoT Based on Parallel CNN</p></title>
  <aug>
    <au><snm>Zeng</snm><fnm>C</fnm></au>
    <au><snm>Wang</snm><fnm>Z</fnm></au>
    <au><snm>Wang</snm><fnm>Z</fnm></au>
  </aug>
  <source>2020 International Conferences on Internet of Things
  (iThings)</source>
  <publisher>Rhodes, Greece: IEEE</publisher>
  <pubdate>2020</pubdate>
  <fpage>258</fpage>
  <lpage>-263</lpage>
</bibl>

<bibl id="B31">
  <title><p>Digital Audio Tampering Detection Based on ENF
  Consistency</p></title>
  <aug>
    <au><snm>Wang</snm><fnm>ZF</fnm></au>
    <au><snm>Wang</snm><fnm>J</fnm></au>
    <au><snm>Zeng</snm><fnm>CY</fnm></au>
    <au><snm>Min</snm><fnm>QS</fnm></au>
    <au><snm>Tian</snm><fnm>Y</fnm></au>
    <au><snm>Zuo</snm><fnm>MZ</fnm></au>
  </aug>
  <source>2018 International Conference on Wavelet Analysis and Pattern
  Recognition (ICWAPR)</source>
  <publisher>Chengdu: IEEE</publisher>
  <pubdate>2018</pubdate>
  <fpage>209</fpage>
  <lpage>-214</lpage>
</bibl>

<bibl id="B32">
  <title><p>Double Compression Detection Based on Feature Fusion</p></title>
  <aug>
    <au><snm>Wang</snm><fnm>ZF</fnm></au>
    <au><snm>Zhu</snm><fnm>L</fnm></au>
    <au><snm>Min</snm><fnm>QS</fnm></au>
    <au><snm>Zeng</snm><fnm>CY</fnm></au>
  </aug>
  <source>2017 International Conference on Machine Learning and Cybernetics
  (ICMLC)</source>
  <publisher>Ningbo: IEEE</publisher>
  <pubdate>2017</pubdate>
  <fpage>379</fpage>
  <lpage>-384</lpage>
</bibl>

<bibl id="B33">
  <title><p>Virtual Chime-Bells Experimental System Based on Multi-Modal
  Fusion</p></title>
  <aug>
    <au><snm>Wang</snm><fnm>Z</fnm></au>
    <au><snm>Liu</snm><fnm>Q</fnm></au>
    <au><snm>Yao</snm><fnm>H</fnm></au>
    <au><snm>Chen</snm><fnm>J</fnm></au>
  </aug>
  <source>2015 International Conference of Educational Innovation through
  Technology (EITT)</source>
  <publisher>Wuhan, China: IEEE</publisher>
  <pubdate>2015</pubdate>
  <fpage>64</fpage>
  <lpage>-67</lpage>
</bibl>

<bibl id="B34">
  <title><p>Digital audio forensics: A first practical evaluation on microphone
  and environment classification</p></title>
  <aug>
    <au><snm>Kraetzer</snm><fnm>C</fnm></au>
    <au><snm>Oermann</snm><fnm>A</fnm></au>
    <au><snm>Dittmann</snm><fnm>J</fnm></au>
    <au><snm>Lang</snm><fnm>A</fnm></au>
  </aug>
  <source>Workshop on Multimedia \& Security</source>
  <pubdate>2007</pubdate>
  <fpage>63</fpage>
  <lpage>74</lpage>
</bibl>

<bibl id="B35">
  <title><p>Source Cell-Phone Identification in the Presence of Additive Noise
  from CQT Domain</p></title>
  <aug>
    <au><snm>Qin</snm><fnm>T</fnm></au>
    <au><snm>Wang</snm><fnm>R</fnm></au>
    <au><snm>Yan</snm><fnm>D</fnm></au>
    <au><snm>Lin</snm><fnm>L</fnm></au>
  </aug>
  <source>Information(Switzerland).</source>
  <pubdate>2018</pubdate>
  <volume>9</volume>
  <fpage>205</fpage>
</bibl>

<bibl id="B36">
  <title><p>Source Digital Voice Recorder Identification by Wavelet
  Analysis</p></title>
  <aug>
    <au><snm>Eskidere</snm><fnm>A</fnm></au>
  </aug>
  <source>International Journal on Artificial Intelligence Tools.</source>
  <pubdate>2016</pubdate>
  <volume>25</volume>
</bibl>

<bibl id="B37">
  <title><p>Optimizing acoustic features for source cell-phone recognition
  using speech signals</p></title>
  <aug>
    <au><snm>Hanilçi</snm><fnm>C</fnm></au>
    <au><snm>Ertas</snm><fnm>F</fnm></au>
  </aug>
  <source>Acm Workshop on Information Hiding \& Multimedia Security</source>
  <pubdate>2013</pubdate>
  <fpage>141</fpage>
  <lpage>148</lpage>
</bibl>

<bibl id="B38">
  <title><p>Cellphone identification using noise estimates from recorded
  audio</p></title>
  <aug>
    <au><snm>{Aggarwal}</snm><fnm>R.</fnm></au>
    <au><snm>{Singh}</snm><fnm>S.</fnm></au>
    <au><snm>{Roul}</snm><fnm>A. K.</fnm></au>
    <au><snm>{Khanna}</snm><fnm>N.</fnm></au>
  </aug>
  <source>2014 International Conference on Communication and Signal
  Processing</source>
  <pubdate>2014</pubdate>
  <fpage>1218</fpage>
  <lpage>1222</lpage>
</bibl>

<bibl id="B39">
  <title><p>Source cell phone matching from speech recordings by sparse
  representation and KISS metric</p></title>
  <aug>
    <au><snm>{Zou}</snm><fnm>L.</fnm></au>
    <au><snm>{He}</snm><fnm>Q.</fnm></au>
    <au><snm>{Yang}</snm><fnm>J.</fnm></au>
    <au><snm>{Li}</snm><fnm>Y.</fnm></au>
  </aug>
  <source>2016 IEEE International Conference on Acoustics, Speech and Signal
  Processing (ICASSP)</source>
  <pubdate>2016</pubdate>
  <fpage>2079</fpage>
  <lpage>2083</lpage>
</bibl>

<bibl id="B40">
  <title><p>Source Smartphone Identification by Exploiting Encoding
  Characteristics of Recorded Speech</p></title>
  <aug>
    <au><snm>Jin</snm><fnm>C</fnm></au>
    <au><snm>Wang</snm><fnm>R</fnm></au>
    <au><snm>Yan</snm><fnm>D</fnm></au>
  </aug>
  <source>Digital Investigation.</source>
  <pubdate>2019</pubdate>
  <volume>29</volume>
  <fpage>129</fpage>
  <lpage>146</lpage>
</bibl>

<bibl id="B41">
  <title><p>Support vector machines using GMM supervectors for speaker
  verification</p></title>
  <aug>
    <au><snm>{Campbell}</snm><fnm>W. M.</fnm></au>
    <au><snm>{Sturim}</snm><fnm>D. E.</fnm></au>
    <au><snm>{Reynolds}</snm><fnm>D. A.</fnm></au>
  </aug>
  <source>IEEE Signal Processing Letters.</source>
  <pubdate>2006</pubdate>
  <volume>13</volume>
  <fpage>308</fpage>
  <lpage>311</lpage>
</bibl>

<bibl id="B42">
  <title><p>Source Microphone Recognition Aided by a Kernel-Based Projection
  Method</p></title>
  <aug>
    <au><snm>{Jiang}</snm><fnm>Y.</fnm></au>
    <au><snm>{Leung}</snm><fnm>F. H. F.</fnm></au>
  </aug>
  <source>IEEE Transactions on Information Forensics and Security.</source>
  <pubdate>2019</pubdate>
  <volume>14</volume>
  <fpage>2875</fpage>
  <lpage>2886</lpage>
</bibl>

<bibl id="B43">
  <title><p>Mobile phone clustering from acquired speech recordings using deep
  Gaussian supervector and spectral clustering</p></title>
  <aug>
    <au><snm>{Li}</snm><fnm>Y.</fnm></au>
    <au><snm>{Zhang}</snm><fnm>X.</fnm></au>
    <au><snm>{Li}</snm><fnm>X.</fnm></au>
    <au><snm>{Feng}</snm><fnm>X.</fnm></au>
    <au><snm>{Yang}</snm><fnm>J.</fnm></au>
    <au><snm>{Chen}</snm><fnm>A.</fnm></au>
    <au><snm>{He}</snm><fnm>Q.</fnm></au>
  </aug>
  <source>2017 IEEE International Conference on Acoustics, Speech and Signal
  Processing (ICASSP)</source>
  <pubdate>2017</pubdate>
  <fpage>2137</fpage>
  <lpage>2141</lpage>
</bibl>

<bibl id="B44">
  <title><p>Subband Aware CNN for Cell-Phone Recognition</p></title>
  <aug>
    <au><snm>{Lin}</snm><fnm>X.</fnm></au>
    <au><snm>{Zhu}</snm><fnm>J.</fnm></au>
    <au><snm>{Chen}</snm><fnm>D.</fnm></au>
  </aug>
  <source>IEEE Signal Processing Letters.</source>
  <pubdate>2020</pubdate>
  <volume>27</volume>
  <fpage>605</fpage>
  <lpage>609</lpage>
</bibl>

<bibl id="B45">
  <title><p>Generalized linear discriminant sequence kernels for speaker
  recognition</p></title>
  <aug>
    <au><snm>{Campbell}</snm><fnm>W. M.</fnm></au>
  </aug>
  <source>2002 IEEE International Conference on Acoustics, Speech, and Signal
  Processing</source>
  <pubdate>2002</pubdate>
  <volume>1</volume>
  <fpage>I</fpage>
  <lpage>161-I-164</lpage>
</bibl>

<bibl id="B46">
  <title><p>Microphone Identification Using Convolutional Neural
  Networks</p></title>
  <aug>
    <au><snm>{Baldini}</snm><fnm>G.</fnm></au>
    <au><snm>{Amerini}</snm><fnm>I.</fnm></au>
    <au><snm>{Gentile}</snm><fnm>C.</fnm></au>
  </aug>
  <source>IEEE Sensors Letters.</source>
  <pubdate>2019</pubdate>
  <volume>3</volume>
  <fpage>1</fpage>
  <lpage>4</lpage>
</bibl>

<bibl id="B47">
  <title><p>A Survey of Techniques for the Identification of Mobile Phones
  Using the Physical Fingerprints of the Built-In Components</p></title>
  <aug>
    <au><snm>{Baldini}</snm><fnm>G.</fnm></au>
    <au><snm>{Steri}</snm><fnm>G.</fnm></au>
  </aug>
  <source>IEEE Communications Surveys Tutorials.</source>
  <pubdate>2017</pubdate>
  <volume>19</volume>
  <fpage>1761</fpage>
  <lpage>1789</lpage>
</bibl>

<bibl id="B48">
  <title><p>Band Energy Difference for Source Attribution in Audio
  Forensics</p></title>
  <aug>
    <au><snm>{Luo}</snm><fnm>D.</fnm></au>
    <au><snm>{Korus}</snm><fnm>P.</fnm></au>
    <au><snm>{Huang}</snm><fnm>J.</fnm></au>
  </aug>
  <source>IEEE Transactions on Information Forensics and Security.</source>
  <pubdate>2018</pubdate>
  <volume>13</volume>
  <fpage>2179</fpage>
  <lpage>2189</lpage>
</bibl>

<bibl id="B49">
  <title><p>Source cell phone verification from speech recordings using sparse
  representation</p></title>
  <aug>
    <au><snm>Zou</snm><fnm>L</fnm></au>
    <au><snm>He</snm><fnm>Q</fnm></au>
    <au><snm>Wu</snm><fnm>J</fnm></au>
  </aug>
  <source>Digital Signal Processing.</source>
  <pubdate>2016</pubdate>
  <volume>62</volume>
  <fpage>125</fpage>
  <lpage>136</lpage>
</bibl>

<bibl id="B50">
  <title><p>Cell phone verification from speech recordings using sparse
  representation</p></title>
  <aug>
    <au><snm>{Zou}</snm><fnm>L.</fnm></au>
    <au><snm>{He}</snm><fnm>Q.</fnm></au>
    <au><snm>{Feng}</snm><fnm>X.</fnm></au>
  </aug>
  <source>2015 IEEE International Conference on Acoustics, Speech and Signal
  Processing (ICASSP)</source>
  <pubdate>2015</pubdate>
  <fpage>1787</fpage>
  <lpage>1791</lpage>
</bibl>

<bibl id="B51">
  <title><p>K-SVD: An algorithm for designing overcomplete dictionaries for
  sparse representation</p></title>
  <aug>
    <au><snm>{Aharon}</snm><fnm>M.</fnm></au>
    <au><snm>{Elad}</snm><fnm>M.</fnm></au>
    <au><snm>{Bruckstein}</snm><fnm>A.</fnm></au>
  </aug>
  <source>IEEE Transactions on Signal Processing.</source>
  <pubdate>2006</pubdate>
  <volume>54</volume>
  <fpage>4311</fpage>
  <lpage>4322</lpage>
</bibl>

<bibl id="B52">
  <title><p>Discriminative K-SVD for dictionary learning in face
  recognition</p></title>
  <aug>
    <au><snm>{Zhang}</snm><fnm>Q.</fnm></au>
    <au><snm>{Li}</snm><fnm>B.</fnm></au>
  </aug>
  <source>2010 IEEE Computer Society Conference on Computer Vision and Pattern
  Recognition</source>
  <pubdate>2010</pubdate>
  <fpage>2691</fpage>
  <lpage>2698</lpage>
</bibl>

<bibl id="B53">
  <title><p>Spatial As Deep: Spatial CNN for Traffic Scene
  Understanding</p></title>
  <aug>
    <au><snm>Pan</snm><fnm>X</fnm></au>
    <au><snm>Shi</snm><fnm>J</fnm></au>
    <au><snm>Luo</snm><fnm>P</fnm></au>
    <au><snm>Wang</snm><fnm>X</fnm></au>
    <au><snm>Tang</snm><fnm>X</fnm></au>
  </aug>
  <pubdate>2017</pubdate>
</bibl>

<bibl id="B54">
  <title><p>Deep Residual Learning for Image Recognition</p></title>
  <aug>
    <au><snm>{He}</snm><fnm>K.</fnm></au>
    <au><snm>{Zhang}</snm><fnm>X.</fnm></au>
    <au><snm>{Ren}</snm><fnm>S.</fnm></au>
    <au><snm>{Sun}</snm><fnm>J.</fnm></au>
  </aug>
  <source>2016 IEEE Conference on Computer Vision and Pattern Recognition
  (CVPR)</source>
  <pubdate>2016</pubdate>
  <fpage>770</fpage>
  <lpage>778</lpage>
</bibl>

<bibl id="B55">
  <title><p>Speech Emotion Classification Using Attention-Based
  LSTM</p></title>
  <aug>
    <au><snm>{Xie}</snm><fnm>Y.</fnm></au>
    <au><snm>{Liang}</snm><fnm>R.</fnm></au>
    <au><snm>{Liang}</snm><fnm>Z.</fnm></au>
    <au><snm>{Huang}</snm><fnm>C.</fnm></au>
    <au><snm>{Zou}</snm><fnm>C.</fnm></au>
    <au><snm>{Schuller}</snm><fnm>B.</fnm></au>
  </aug>
  <source>IEEE/ACM Transactions on Audio, Speech, and Language
  Processing.</source>
  <pubdate>2019</pubdate>
  <volume>27</volume>
  <fpage>1675</fpage>
  <lpage>1685</lpage>
</bibl>

<bibl id="B56">
  <title><p>Support vector machines using GMM supervectors for speaker
  verification</p></title>
  <aug>
    <au><snm>Campbell</snm><fnm>W.M.</fnm></au>
    <au><snm>Sturim</snm><fnm>D.E.</fnm></au>
    <au><snm>Reynolds</snm><fnm>D.A.</fnm></au>
  </aug>
  <source>IEEE Signal Processing Letters</source>
  <pubdate>2006</pubdate>
  <volume>13</volume>
  <issue>5</issue>
  <fpage>308</fpage>
  <lpage>311</lpage>
</bibl>

</refgrp>
} 







\end{backmatter}
\end{document}